\documentclass{aa} 
\usepackage{booktabs}
\usepackage{graphicx}
\usepackage{ gensymb }
\usepackage{ textcomp }
\usepackage{ wasysym }
\usepackage{float}
\usepackage{rotating}
\usepackage{lscape}
\begin{document} 
\titlerunning{On water delivery in the inner solar nebula}
\authorrunning{D'Angelo et al.}
\title{On water delivery in the inner solar nebula}
  \subtitle{Monte Carlo simulations of forsterite hydration}
 
 \author {
 	M. D'Angelo \inst{1,2},
	S. Cazaux \inst{3,4},  
  	I. Kamp \inst{2}, 
        W.- F. Thi \inst{5}  
\and P. Woitke \inst{6}
         }
 \institute{Zernike Institute for Advanced Materials, University of Groningen, P.O.\ Box 221, 9700 AE Groningen, The Netherlands, \email{dangelomartinka@gmail.com}
         \and
          Kapteyn Astronomical Institute, University of Groningen, P.O.\ Box 800, 9747 AV Groningen, The Netherlands, \email{kamp@astro.rug.nl}
             \and
            Faculty of Aerospace Engineering, Delft University of Technology, Delft, The Netherlands
           \and
            University of Leiden, P.O.\ Box 9513, NL, 2300 RA, Leiden, The Netherlands
         \and
         Max Planck Institute for Extraterrestrial Physics, Gie\ss enbachstrasse 1, 85741 Garching, Germany 
          \and
         SUPA, School of Physics \& Astronomy, University of St.\ Andrews, North Haugh, St.\ Andrews KY16 9SS, UK
                      }
\abstract
{Endogenous or exogenous, dry or wet, various scenarios have been so far depicted for the origin of water on our Solar System's rocky bodies. Hydrated silicates found in meteorites and in interplanetary dust particles together with observations of abundant water reservoirs in the habitable zone of protoplanetary disks are evidences that support aqueous alteration of silicate dust grains by water vapor condensation in a nebular setting.}
   {We investigate the thermodynamics (temperature and pressure dependencies) and kinetics (adsorption rates and energies, surface diffusion and cluster formation) of water adsorption on surfaces of forsterite grains, constraining the location in the solar nebula where aqueous alteration of silicates by water vapor adsorption could occur efficiently and lead to the formation of phyllosilicates. We analyze the astrophysical conditions favorable for such hydration mechanism and the implications for water on solid bodies.}
   {The protoplanetary disk model (ProDiMo) code is tuned to simulate the thermochemical disk structure of the early solar nebula at three evolutionary stages. Pressure, temperature and water vapor abundance within $1$~au from the protosun were extracted and used as input for a Monte Carlo code to model water associative adsorption using adsorption energies that resemble the forsterite [100] crystal lattice.}
   {Hydration of forsterite surfaces by water vapor adsorption could have occurred within the nebula lifetime already at a density of $10^8$~cm$^{-3}$, with increasing surface coverage for higher water vapor densities. Full surface coverage is attained for temperatures lower than $500$~K, while for hotter grain surfaces water cluster formation plays a crucial role. Between $0.5$ and $10$ number of Earth's oceans can arise from the agglomeration of hydrated $0.1$~$\mu$m grains into an Earth-sized planet. However, if grain growth occurs dry and water vapor processes the grains afterwards, this value can decrease by two orders of magnitude.}
   {This work shows that water cluster formation enhances the water surface coverage and enables a stable water layer to form at high temperature and low water vapor density conditions. Finally, surface diffusion of physisorbed water molecules shortens the timescale for reaching steady state, enabling phyllosilicate formation within the solar nebula timescale.}

\maketitle

\section{Introduction}\label{intro}
After almost $40$ years of study, the origin of Earth's water is still strongly debated \citep{Drake2005}. 
One hypothesis is that Earth accreted from a mixture of dry and wet primary building blocks in which water was in the form of hydrous silicates (the \emph{wet-endogenous} scenario); another view supports dry accretion, with water delivered at a later stage during impact of hydrous asteroidal or cometary bodies (the \emph{exogenous} scenario).

A direct evidence of aqueous alteration processes in the solar nebula is contained in Carbonaceous Chondrites (CCs). These "undifferentiated" meteorites are considered primitive Solar System objects along with Interplanetary Dust Particles (IDPs) and cometary grains because they show solar composition~\citep{Barrat2012}. Depending on the chemical and mineralogical composition and size of their parent body, several classes of CCs are defined \citep{Weisberg2006}. Among them, CI (Ivuna-like) group, CM (Mighei-like) group and CR (Renazzo-like) carbonaceous chondrites are the most hydrous varieties, with $3$ to $14$~wt\% of water content in CM and CR, and up to $15$~wt\% in CI~\citep{Alexander2010}. 
Most of the water in these chondrites is structurally bound in phyllosilicates\footnote{Layered silicate platelets with swelling properties~\citep{Schuttlefield2007}.} that formed during aqueous alteration of anhydrous minerals (e.g. olivine and pyroxene) very likely on the meteorites parent bodies~\citep{Brearley2006}. Recent mid-IR spectroscopy measurements revealed that the most aqueous altered samples are ($-$OH)-rich and almost depleted in olivine, the "dry" precursor mineral~\citep{Beck2014}.

The high variability in the abundance and in the nature of these hydro-silicates in CCs indicates many levels of aqueous alteration and suggests different possible origins and evolution. Most of the models that have been developed over the last $30$ years are based on the fluid flow and liquid water-rock interaction on their parent bodies: previously accreted water ice melts, the fluid flows through various mineral matrices of different permeability and reacts with the anhydrous precursor mineral finally forming the hydrated products~\citep[see review by][]{Brearley2006}. The chemical composition of phyllosilicates forming by the flow of fluids may be controlled by the composition of the anhydrous precursor mineral and/or the composition of the aqueous solutions~\citep{Howard2011,Velbel2012}.

On the other hand, hydrated silicates could have formed by direct condensation of water vapor within the terrestrial planets forming region. First models indicated that silicate hydration would be kinetically inhibited in a nebular setting~\citep{Fegleyprinn1989}. Using a Simple Collision Theory (SCT) model and the activation energy of $8420$~K ($70$~kJ mol$^{-1}$) as the amount of energy required to convert MgO into Mg$\mathrm{(OH)}_2$ (brucite) at $1$~atmosphere, they estimated the formation rates of serpentine and brucite and concluded that formation of hydrous silicates takes too long to occur by nebular condensation ($\mathrm{10}^5$ times the nebular life time of~$10^{13}$~s). However,~\citet{Ganguly1995} used the same SCT approach and estimated a shorter time scale for the hydration of olivine if a lower activation energy (about $3909$~K, that is $32.5$~kJ mol$^{-1}$) is assumed in the calculation. 

In the attempt to explain the presence of phyllosilicates fine-grained rims (FGRs) in the Murray CM chondrite,~\citet{Ciesla2003} draw a scenario in which, holding the  $8420$~K of hydration energy, shock waves pass through an icy region of the nebula, the water vapor partial pressure is locally enhanced, thus increasing the collision rates of water molecules with the bare grains. Therefore, hydrated silicates can form much faster than the solar nebula life time, allowing a nebular origin of the chondrules and the phyllosilicates components as well.~\citet{Woitke2017} consider phyllosilicates in thermo-chemical equilibrium, and found that below $345$~K and at one bar, the dominant phyllosilicate is $\mathrm{Mg}_3\mathrm{Si}_2\mathrm{O}_9\mathrm{H}_4$ (lizardite) which replaces $\mathrm{Mg}_2\mathrm{SiO}_4$ in chemical + phase equilibrium.

Phyllosilicates can retain water when heated up to a few hundreds of degrees centigrade~\citep{Beck2014, Davies1996} being able to preserve structural water also in the inner and warmer regions of a protoplanetary disk. Once agglomerated into planetesimals, phyllosilicates could be a potential source of water for terrestrial planets, in line with the wet-endogenous scenario.

More recent computer simulations have studied water adsorption energy, binding sites and mechanisms (associative and/or dissociative) on forsterite surfaces and demonstrated that many Earth oceans could efficiently form \emph{in situ} under accretion disk conditions~\citep{Stimpfl2006, Muralidharan2008, King2010, Asaduzzaman2013, Asaduzzaman2015, Prigiobbe2013}. However, these modeling attempts possess some major uncertainties, namely a detailed temperature-pressure structure of the young solar nebula. 
In the exploratory rate-based warm surface chemistry model of~\citet[][submitted]{Thi2018}, water from the gas-phase can chemisorb on
dust grain surfaces and subsequently diffuse into the silicate bulk. The phyllosilicate formation model was applied to a zero-dimensional
chemical model and to a 2D protoplanetary disk model (ProDiMo) to investigate the formation of phyllosilicates in protoplanetary disks.

In this work we test the possibility of water vapor condensation on bare forsterite grains in the region of the terrestrial planets prior to their accretion into planetesimals. In the endogenous scenario, we want to quantify how much water could have been delivered to planetesimal precursors of Venus, Earth and Mars $4.5$~Gyr ago. We used the astrophysical model for protoplanetary disks, ProDiMo~\citep{Woitke2009}, and the Monte Carlo (MC) simulation optimized for studying accretion of ice mantles on grains~\citep{Cazaux2010, Cazaux2015}, both described in sections~\ref{prodimoinputs} and~\ref{MCinput}. Using T Tauri disks observed in the Orion Nebula as templates, with ProDiMo we carefully build up our early solar nebula model at three time steps in the Sun's evolution. Temperature and pressure radial profiles and water vapor abundance are then extracted specifically for the midplane region close to the protosun. We use the MC simulations to calculate water adsorption rates. Surface coverages at different physical conditions were then estimated and used to quantify possible scenarios on the origin of water on terrestrial planets and meteorites.
 \begin{table}
        \centering
         \caption[]{List of the stellar and disk parameters}
      \label{parameter}          
    \begin{tabular}{l l l}
            \hline\hline         
            \noalign{\smallskip}
            Parameters & Symbol  & Value\\
            \noalign{\smallskip}
            \cline{1-3}
            \noalign{\smallskip}
           Stellar Mass         &  $M_{*}$  & $1$~M$_{\odot}$     \\
           Stellar Luminosity & $ L_{*}$    &  $11.02$,\,$2.17$,\,$0.46$\,L$_{\odot}$   \\
 	  Effective temperature   &  $T_{\rm eff}$ & $4147$, $4282$, $4290$~K    \\
           UV luminosity\textsuperscript{a}   &$L_{\rm UV}$  &$0.01$~L$_{\odot}$ \\
           X-ray luminosity\textsuperscript{a}&$L_{\rm X}$ & $10^{30}$~erg s$^{-1}$   \\
           \scalebox{0.95}[1]{Cosmic ray ionization rate} & CRI & $1.7~\times~10^{-17}$~s$^{-1}$\\
            \noalign{\smallskip}
            \cline{1-3}
            \noalign{\smallskip}
        Disk mass & $M_{\rm d}$&  $0.003$, $0.03$~M$_{\odot}$\\
        Disk inner radius & $R_{\rm in}$ &  $0.07$~au \\  
        Disk outer radius  & $R_{\rm out}$   & $100$~au\\ 
        Tapering-off radius   & $R_{\rm tap}$ & $50$~au\\
        Reference scale height & $H_ 0$ & $0.4$~au\\
        Reference radius & $R_{\rm ref}$ & $10$~au\\
         \noalign{\smallskip}
            \cline{1-3}
            \noalign{\smallskip}
     Dust settling turbulence & $\alpha$ & $0.1$ \\
     Column density index\textsuperscript{a}  &  $\epsilon$& $1.0$\\
     Dust-to-gas mass ratio\textsuperscript{a}  & $\rho_{\rm d}/\rho_{\rm g}$  & $0.01$ \\
        \noalign{\smallskip}
            \cline{1-3}
            \noalign{\smallskip}
       Min. size dust grain\textsuperscript{a}  & a$_{\rm min}$ & $0.05~\mu$m\\
       Max. size dust grain\textsuperscript{a}  & a$_{\rm max}$ & $3000~\mu$m\\
       Dust size distr. index\textsuperscript{a}  & $p$ & $3.5$\\
       Dust composition\textsuperscript{a}:  &  &\\ 
      $\mathrm{Mg}_{0.7}\mathrm{Fe}_{0.3}$Si$\mathrm{O}_3$ && 60\%\\
        amorph. carbon  &  &$15$\%\\
        porosity  &  &$25$\%\\
         Dust material density& $\rho_{\rm gr}$ & $2.076$~g \mbox{cm$^{-3}$}\\
          \noalign{\smallskip}
            \cline{1-3}
         \end{tabular}
\tablefoot{ProDiMo input parameters used to model six early solar nebula, for two disk mass values ($0.003$~M$_{\odot}$ and $0.03$~M$_{\odot}$) at three nebular ages ($0.2$~Myr, $1$~Myr and $10$~Myr), to which the following stellar parameters ($L_{*}$ and $T_{\rm eff}$) 
correspond respectively ($11.02$~L$_{\odot}$,~$4147~{\rm K}$), ($2.17$~L$_{\odot}$,~$4282~{\rm K}$) and ($0.46$~L$_{\odot}$,~$4290~{\rm K}$).\\
\tablefoottext{a}{Standard values from~\citet{Helling2014} and~\citet{Woitke2009}.}
} 
\end{table}

\begin{figure*}[h!tp]
\centering
{\includegraphics [width=\columnwidth]{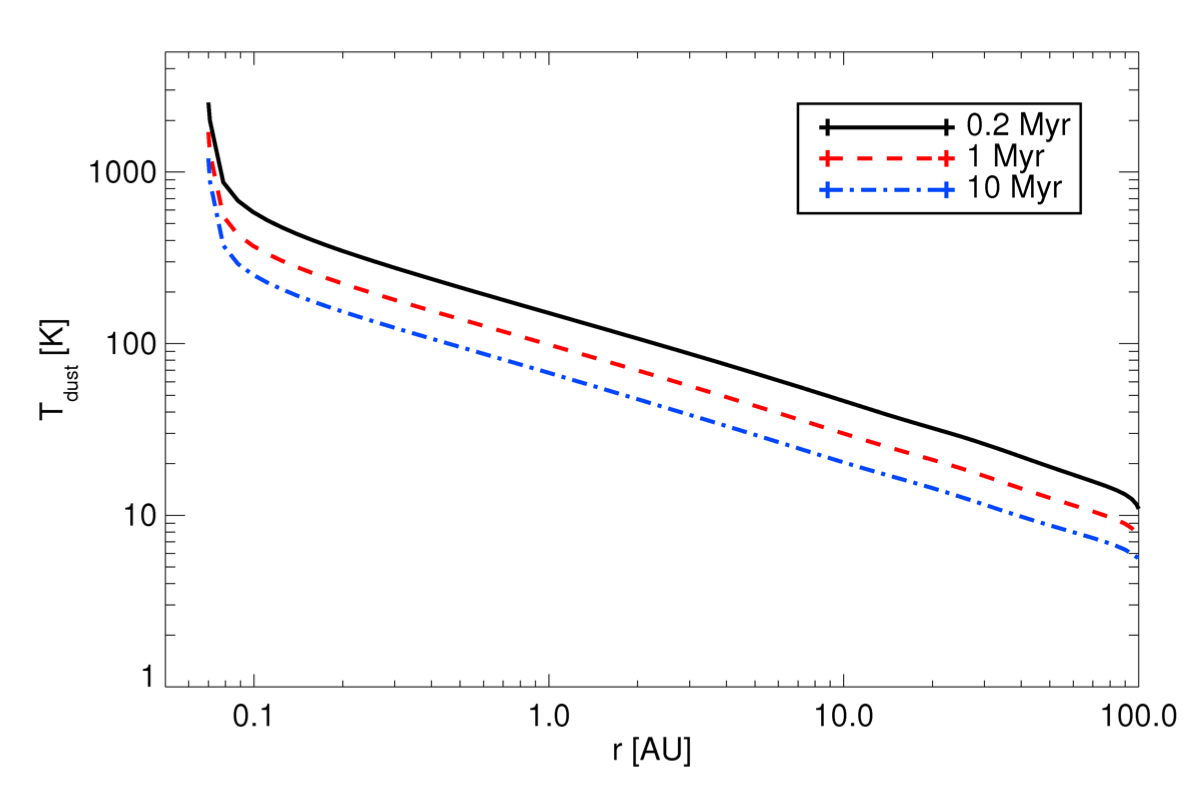}}
{\includegraphics [width=\columnwidth]{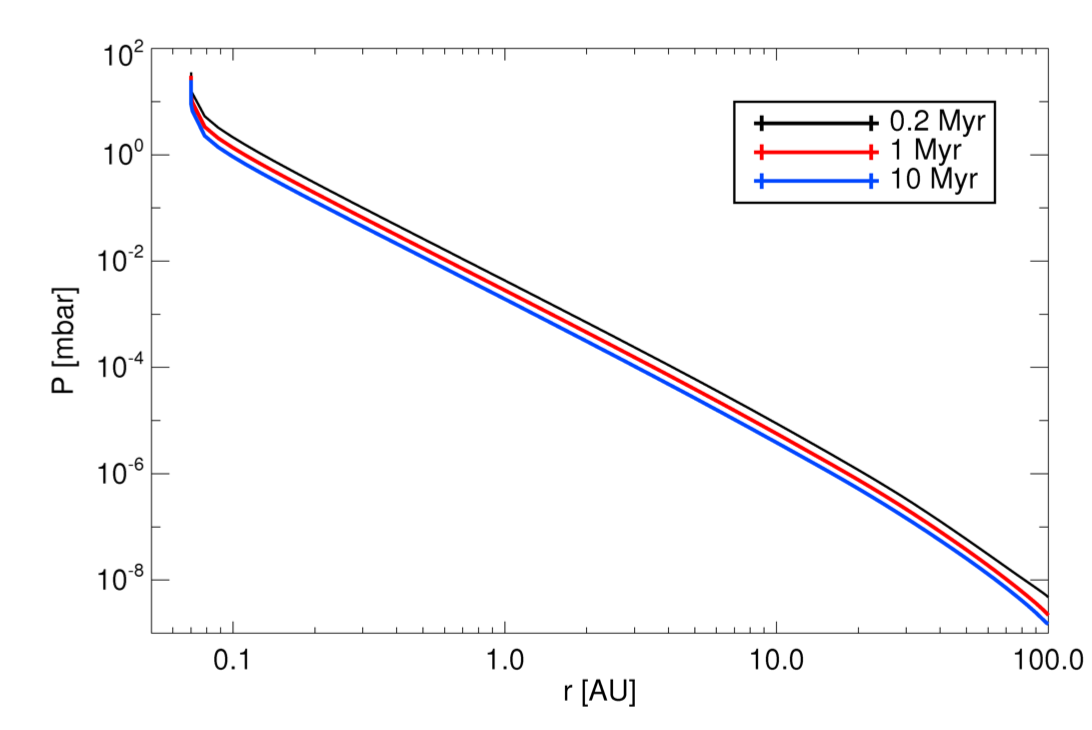}}
\caption{{Midplane ($z/r=0.00$) dust temperature (left) and pressure (right) radial profiles extracted from ProDiMo models of the early solar nebula at three ages ($0.2$~Myr, $1$~Myr and $10$~Myr) and  $0.03$~M$_{\odot}$ disk mass. 
}}
\label{fig:PTprofiles}
\end{figure*}

\section{Simulating the early solar nebula with ProDiMo}\label{prodimo}

Over the last decades the thermo-chemical evolution of our solar nebula has been modeled with different approaches and assumptions. In some of them, the temperature and pressure gradients at different nebula stages were extrapolated from the condensation curves for actual solar system's bodies~\citep{Cameron1995, Fegley1999, Lewis1974}, and in others the temperature and pressure gradients are a result of simulations of protoplanetary disks with typical solar nebula parameters~\citep{Willacy1998, Makalkin2009}. 

To this latter approach belongs the (1+1D) model developed by~\citet{Davis2005a, Davis2005b} to study the dependency of the surface density radial distribution from the disk accretion rate.
~\citet{Min2011} modeled the solar nebula conditions using~$3$D radiative transfer to investigate how the snow line changes with evolving mass accretion rate.~\citet{Hersant2001} used the (1+1D) disk structure turbulent model of~\citep{Hure2001} applied to the solar nebula, to extract the density and temperature profiles favorable for the D/H enrichment in LL3 meteorites and comets.~\citet{Albertsson2014} combined an extended gas-grain chemical model, which accounts for high-temperature and surface reactions with multiply deuterated species, with a (1+1D) steady-state $\alpha$-viscosity nebular model to obtain molecular abundances and D/H ratios for a $1$~Myr old solar nebula.
However, in all these models, the input parameters are not necessarily informed by typical protoplanetary disks as they are observed now in tremendous detail.

The early solar nebula disk structure presented in this work uses the protoplanetary disk modeling code ProDiMo~\citep{Woitke2009, Aresu2012}. The code was developed to consistently calculate the physical, thermal and chemical structure of protoplanetary disks. It uses $2$D dust continuum radiative transfer, gas phase and photo-chemistry and a detailed thermal energy balance for the gas. ProDiMo models have been successfully applied to explain simultaneously multi-wavelength observations of dust and gas (including resolved images) in disks around young stars~\citep[see][]{Thi2010, Woitke2011, Tilling2012, Garufi2014}. 
The code uses now the more realistic disk dust opacities from~\citet{Min2016}, which can simultaneously reproduce thermal, scattering and polarization data from disks. Recently,~\citet{Woitke2016} proposed a parametrized set-up for disk models that can capture enough complexity to match observations without introducing too many free parameters. 

In the following, we use this parametrized set-up of ProDiMo to simulate a young solar nebula under steady-state condition around a Sun-like star at three evolutionary stages. 

\subsection{Physical input parameters}\label{prodimoinputs}

The study of the oxygen isotope fractionation found in meteoritic mineral inclusions revealed that the protosun probably formed in a high mass star forming region at a distance of $\sim\!1$~parsec from an O or B star~\citep{Young2011}.  However, some uncertainty remains as to the cluster size \citep{Adams2010}. Therefore, protoplanetary disks observed in the Orion Nebula (also called proplyds) are used here as templates to build our astrophysical model of the early solar nebula. The stellar and disk input parameters chosen for this work are listed in Table~\ref{parameter}. 

UV and X-ray luminosities are fixed to standard values in circumstellar disks~\citep{Woitke2016}. We neglect here the presence of the external UV radiation field by the possible nearby O and B stars; its impact on the disk midplane temperatures inside $\sim\!1$~au --- the region relevant for our study of dust hydration --- is negligible \citep{Walsh2013}. However, as shown by e.g.\ \citet{Walsh2013}, \citet{Antonellini2015} and \citet{Rab2018} such enhanced external UV and X-ray radiation fields can have profound consequences for disk surface layers and the outer disk midplane which are readily observable through mid-IR and submm line emission of water and ion molecules.

Throughout the solar nebula, dust abundance and size distribution are assumed constant, with the latter following the power-law $f(a)\!\sim\!a^{-p}$, with index $p\!=\!3.5$~\citep{Woitke2009} and grain sizes between $a_{\rm min}\!=\!0.05~\mu$m and $a_{\rm max}\!=\!~\mu$m (see Table~\ref{parameter}). The dust in our models is that part of solids that are accessible through observations of protoplanetary disks such as SEDs. These disks could contain already larger solids, but evidence for that is so far indirect from e.g.\ dating meteorites and putting constraints on ages of their parent asteroids in our Solar System \citep[e.g.][]{Amelin2005,Amelin2006} or disk substructure as revealed by SPHERE and ALMA images \citep[e.g.][]{Perez2014,Benisty2015} which could indicate planetary mass companions \citep{deJuanOverlar2016}.

The comparison between the disk mass distribution of the SubMillimeter Array (SMA) survey of $55$ proplyds in Orion and similarly-aged disks in the low mass star forming regions Taurus and Ophiuchus shows that the Orion disk distribution is statistically different from the other two. The number of disks per logarithmic mass bin is approximately constant for masses $0.004 - 0.04$~M$_{\odot}$ in all three regions, but Orion lacks disks more massive than $0.04$~M$_{\odot}$~\citep{Mann2012}. Accordingly, $0.03$~M$_{\odot}$ and $0.003$~M$_{\odot}$ were chosen as representative values for the disk mass of our early solar nebula.

Most of the previous solar nebula models are based on the Hayashi Minimum-Mass-Solar Nebula representation (MMSN), where the local surface density is given by the mass of each planet spread on an appropriate annular area. In that case the surface density scales as a power law with index equal to $-3/2$~\citep{Weidenschilling1977, Hayashi1981}. The surface density profile in our work is assumed to be a power law
\begin{equation}
\Sigma(r) = \Sigma_{0} r^{-\epsilon} e^{(-r/R_{\rm tap})},
\label{eq:sigma}
\end{equation} 
with index $\epsilon$ equal to $1$, less steep than the MMSN but in agreement with observations of proplyds in Orion Nebula~\citep{Mann2010}. The exponential factor causes a tapering-off for the outer edge, meaning that  at $R_{\rm tap}$ the disk surface density profile starts an exponential cut-off and most of the disk mass will be contained therein.

The disk size was extracted from the disk diameter distribution histogram made for the total sample of $149$ proplyds observed in the Trapezium cluster of Orion Nebula with the HST~\citep{Vicente2005} and the SMA~\citep{Mann2010}. It indicates that $75$ to $80$\% of disks have diameters smaller than $150$~au and $40$\% of those have disk radii larger than $50$~au. Hence, we picked a radius of $50$~au as taper radius for our solar nebula.

In ProDiMo, the vertical disk structure is fully parametrized. Given a scale height, $H_0$, at a reference radius, $R_{\rm ref}$, the scale height of the disk is given as
\begin{equation}
H= H_0 {(r/R_{\rm ref})}^ {1.1}.
\label{eq:H}
\end{equation}

This work studies a disk at the end of the cloud core collapse, corresponding to a protostar~$+$~disk system older than $10^{5}$~years. Three evolutionary stages are considered here: $0.2$~Myr, $1$~Myr and $10$~Myr. The dust temperature profile is given by solving the $2$D dust continuum radiative transfer equation. The luminosity and effective temperature of our protosun were picked from the evolutionary tracks that the Grenoble stellar evolution code for pre-main sequence stars~\citep{Siess2000} gives for a $1$~M$_{\odot}$ star of solar metallicity ($Z\!=\!0.02$) at three evolutionary ages: $0.2$~Myr, $1$~Myr and $10$~Myr. We find stellar luminosity ($L_{*}$) and effective temperature ($T_{\rm eff}$) pairs respectively of ($11.02$~L$_{\odot}$,~$4147~{\rm K}$), ($2.17$~L$_{\odot}$,~$4282~{\rm K}$) and ($0.46$~L$_{\odot}$,~$4290~{\rm K}$). We use here passive disk models, i.e.\ we neglect the extra heating in the inner disk midplane regions that stems from accretion. Even though we observe mass accretion through e.g.\ studies of emission lines, the underlying momentum transport and how the mass accretion occurs is less clear than it was in the past; this is due to simulations now including non-ideal MHD effects \citep[e.g.][]{Lesur2014} and recent ALMA studies putting strong limits on levels of turbulence, both indicating that the disks are in large parts more laminar than originally thought \citep[e.g.][]{Flaherty2018}. 

 \subsection{Temperature and pressure disk radial profiles}\label{prodimoresults}

The dust temperature ($T_{\rm dust}$) and pressure radial profiles at the disk midplane ($z/r= 0.00$) were extracted from six early solar nebula models: two disk mass limits ($0.003$~M$_{\odot}$ and $0.03$~M$_{\odot}$) at three nebular ages ($0.2$~Myr, $1$~Myr and $10$~Myr). In the following figures, only the values for the highest disk mass are shown for simplicity. 

The youngest disk at the inner radius (fixed to $0.07$~au) is hotter than the other two older models (Fig.~\ref{fig:PTprofiles}, left). The main difference in the three ages is the stellar luminosity, which determines the main heating source in the innermost regions close to the protosun. The dust inner-rim is heated by the stellar radiation, which is stronger for the youngest protosun. The maximum temperature far exceeds the dust condensation temperature. It ranges from about $2700$~K for the $0.2$~Myr old disk (black curve) to $1770$~K at $1$~Myr (red curve) and finally to about $1210$~K for the $10$~Myr model (blue curve). Here the inner radius was not adapted to a unified dust condensation temperature, since our primary goal is not to capture the intricate details of the inner rim of the disk. Our study focuses on the midplane region where $T_{\rm dust}$ ranges between $300$~K and $600$~K (Sec.~\ref{MCinput}). There are no differences in the midplane temperature profiles for the two disk masses considered.

The radial pressure profile (right side of Fig.~\ref{fig:PTprofiles}) follows the trend of the dust temperature. Indeed, from Eq.~(\ref{eq:sigma}) and Eq.~(\ref{eq:H}), it is clear that the column density is fixed and the mass is distributed according to our prescription, so the volume density is not changing with age in our models.
For disks with mass $0.03$~M$_{\odot}$, the pressure is simply one order of magnitude higher than for $0.003$~M$_{\odot}$ disks. This is true for the very optically thick part of the midplane.

\begin{figure}[htp]
\vspace*{-3mm}
\centering
{\includegraphics [width=1.\columnwidth]{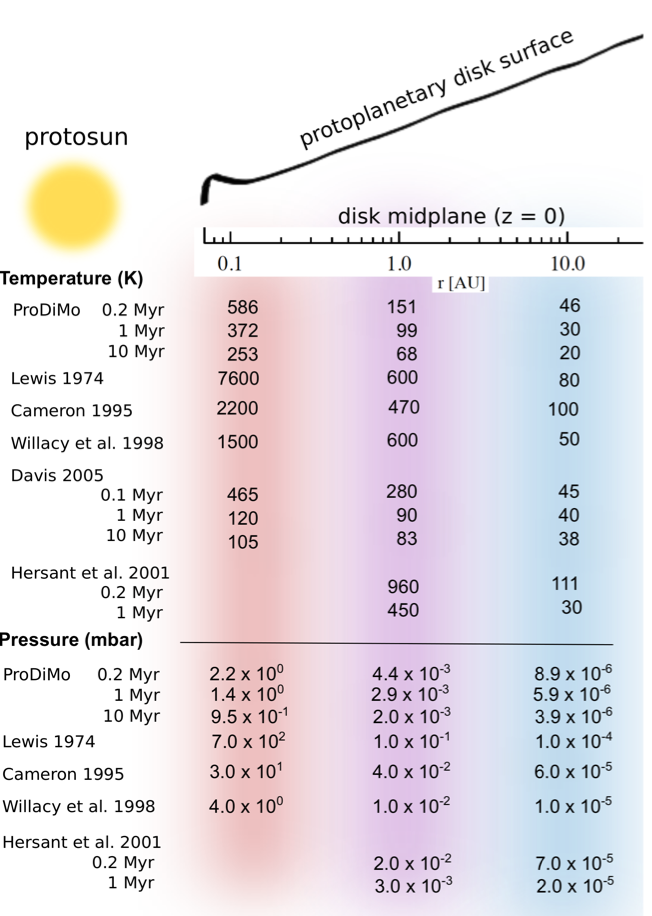}}\\
\caption{{Overview of the thermal and pressure profiles of our early solar nebula model compared to previous simulations. ProDiMo values are for a $0.03$~M$_{\odot}$ disk. Values from~\citet{Lewis1974},~\citet{Cameron1995} and~\citet{Willacy1998} used in Fig.~1 of~\citet{Fegley1999} are extrapolated to $0.1$~au. ($P-T$) values of~\citet{Hersant2001} and~\citet{Davis2005a} are extracted from their figures.}}
\label{fig:Sketch}
\end{figure}

Figure~\ref{fig:Sketch} offers an overview of the state-of-the-art of previous solar nebula simulations in terms of temperature and pressure values at three representative distances from the protosun. A big discrepancy between our temperature values and the ones given in~\citet{Lewis1974},~\citet{Cameron1995} and~\citet{Willacy1998} simulations exists throughout the disk. At $0.1$~au, our youngest nebula ($0.2$~Myr old) is two times colder than~\citet{Willacy1998} nebula and a factor of $13$ colder than~\citet{Lewis1974} model. At $1$~au the gap decreases to a factor of four and at $10$~au our disk becomes two times colder indicating a very different slope of $T(r)$. For the pressure, there is an overall good agreement, except for the values in~\citet{Lewis1974} model which are two orders of magnitude larger than ours. ~\citet{Lewis1974} and~\citet{Cameron1995} models, based on the condensation curves of the actual composition of the solar system's bodies, predict an adiabatic temperature-pressure dependency, far from our ProDiMo disks in thermal equilibrium. The $1$D vertical structure model of a viscously heated disk ($\alpha=0.01$, $\dot M =10^{-7}$~M$_{\odot}$~yr$^{-1}$, $R_{\rm disk}\!=\,$[$0.1-100$]~au) described by~\citet{Willacy1998} is slightly closer to ours. However, differences in input parameters likely cause the temperature differences. Differences are also seen for the ($P-T$) values taken from the solar nebula simulation of~\citet{Hersant2001}. The model was calculated using $\alpha\!=\!0.009$, $M_{\rm disk}\!=\!0.3$~M$_{\odot}$ and it results in $R_{\rm out}\!=\!42$ and $32$~au respectively for $1$ and $0.2$~Myr old disk.
A good agreement exists between our temperature values and those of~\citet{Davis2005a}, who implemented a $\beta$-prescription for the viscous heating in the 2D disk model from~\citet{Dullemond2002}.

In an active disk the viscous heating of the gas by accretion of material from the disk towards the protosun increases the dust temperature in the midplane by thermal accomodation with the gas. The effect of viscous heating is not captured in our ProDiMo models of a passive disk, in which the column density is fixed, the mass is distributed according to our prescription and dust and gas are thermally coupled in the midplane. This can explain the large deviations between the accretion models discussed earlier and the set of simulations studied in this work.
\begin{figure*}[htp]
\centering
{\includegraphics [width=2.0\columnwidth]{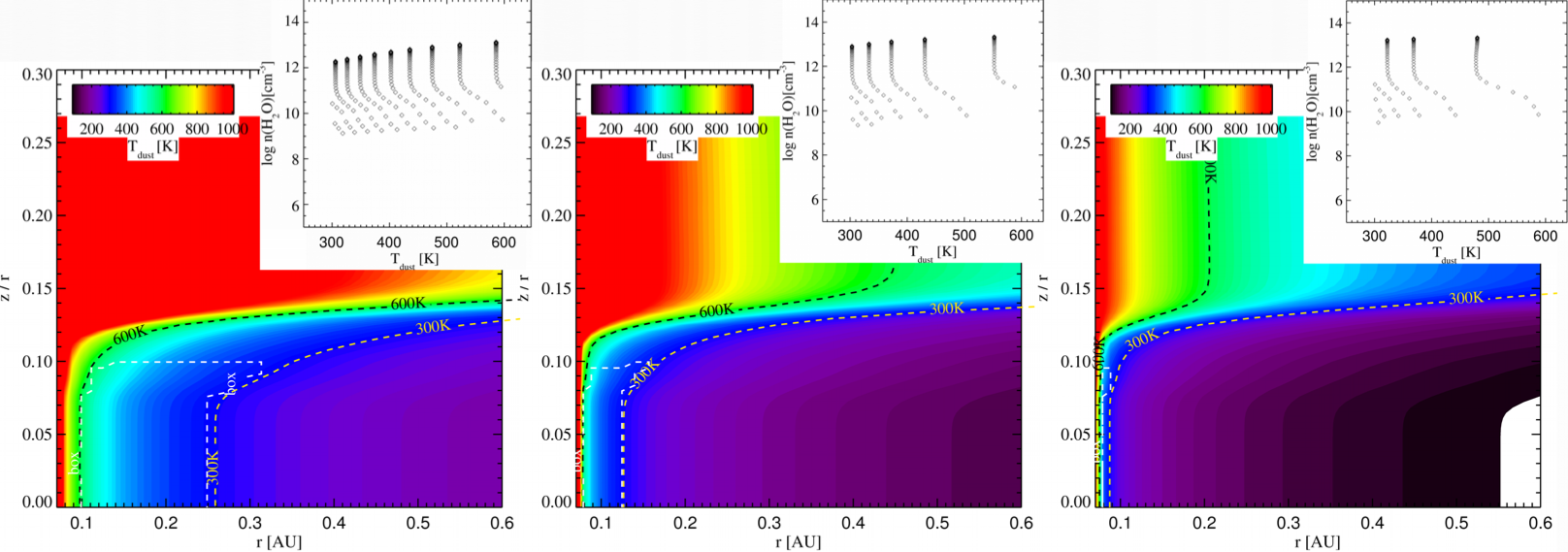}}
\caption{{From left to right: dust temperature structure $T_{\rm dust}$($r$, $z/r$) from ProDiMo models for a $0.03$~M$_{\odot}$ disk respectively $0.2$~Myr, $1$~Myr and $10$~Myr old. White dashed lines identify the ($T, n_{\rm H_2O}$) parameters space called \emph {box} and analyzed in this work. The insert figures show the logarithmic scale of water vapor density $n_{\rm H_2O}$ as a function of $T_{\rm dust}$. Within the \emph {box} it ranges between $10^{9}$ and $10^{13}$~cm$^{-3}$.}}
\label{fig:combi}
\end{figure*}
\section{Monte Carlo simulation of water adsorption on forsterite surface}\label{MC}

In order to test the wet endogenous scenario,~\citet{Stimpfl2004} quantified the amount of water molecules adsorbed on grain surfaces. A grid of $10\,000$ sites, considered to mimic the dust surface, could interact with an infinite reservoir of water molecules, where a maximum of one monolayer is allowed for physisorption with the bare surface (adsorption energy about $600$~K, corresponding to $5$~kJ~mol$^{-1}$). They consider the increase of binding energy due to the cooperative behavior of neighboring water molecules for the formation of clusters. At steady state only $0.25$\% of Earth's ocean could accrete at $1000$~K and $3$\% of it at $500$~K.\\
~\citet{Stimpfl2006} considered an atomistic approach to investigate water adsorption kinetics on the [100] and [010] forsterite crystal planes. The calculation of the surface energy potential distribution showed that the forsterite surface is mostly hydrophobic since the low coordinated surface Mg atoms are the only "attractive" binding sites.

Based on this work,~\citet{Muralidharan2008} studied the mechanisms of adsorption of water onto forsterite surfaces combining an energy minimization technique with a kinetic Monte Carlo simulation. This study showed that at $700$~K and $10^{-5}$~mbar of water partial pressure a single layer coverage of $15$~H$_2$O~nm$^{-2}$ forms for the [100] crystal lattice, while the [010] orientation is less reactive (about $1$~H$_2$O~nm$^{-2}$).

\citet{King2010} investigated the chemisorption of water molecules with adsorption energies as high as $38\,511.5$~K ($320.2$~kJ mol$^{-1}$) versus $19\,243.7$~K ($160.0$~kJ mol$^{-1}$) of~\citet{Stimpfl2006} for the stoichiometric [100] lattice. Their calculation also showed that defective [100] and [010] surfaces are more reactive and hence favorable to H$_2$O adsorption than the stoichiometric ones, playing a crucial role at lower water partial pressures ($\sim\!10^{-8}$~bar) and high temperatures ($\sim\!1500$~K). Subsequent numerical calculations extended the investigation from vacuum to ambient conditions~\citep{Prigiobbe2013} and included different crystal orientations and adsorptions sites of the mineral~\citep{Asaduzzaman2013, Asaduzzaman2015}. However, a detailed temperature-pressure structure of the young solar nebula is needed as a benchmark to evaluate the wider astrophysical implications.

In this work we have investigated water adsorption by means of a Monte Carlo (MC) numerical code developed by~\citet{Cazaux2010, Cazaux2015}. With our simulations we address the following questions:
\begin{enumerate}
\item How much water molecules can adsorb on dust surfaces according to the ($T, n_{\rm H_2O}$) parameters space typical from morphological (thermal and aqueous) alterations of the grains?
\item Which surface mechanisms and properties (adsorption, evaporation, binding energy, cluster formation, etc.) compete for the formation of the first water layer?
\item Where in the nebula can water vapor condensation efficiently hydrate meteoritic and asteroidal mineral components?
\item Is this a possible scenario to explain the presence of water on Earth?
\end{enumerate}

\subsection{Input parameters}\label{MCinput}

Our study focuses on the hydration of forsterite surface grains by water vapor condensation in the habitable zone. We have defined a region in the disk midplane ($z/r$ < $0.1$) where the gas and dust temperatures are coupled and referred to as the surface temperature in our MC models ($T_{\rm gas}\!=\!T_{\rm dust}\!=\!T$) and range from $300$~K up to $600$~K. This \emph {box} changes location in the nebula, moving inward or outward and/or shrinking depending on the nebula's age, hence stellar luminosity (see Fig.~\ref{fig:combi}). The H$_2$O vapor density as function of temperature (inserts of Fig.~\ref{fig:combi}), which correspond to the conditions in the box, was extracted from each ProDiMo model. Three values were here used as input parameters in the MC simulation within the range [$10^{8} - 10^{13}$]~cm$^{-3}$ (see Table~\ref{tab:MCparameter}).

In our MC simulation the [100] forsterite crystal lattice is considered and consists of a grid composed by $20 \times 20$ sites with a total surface area of $27.04$~nm$^2$. According to previous DFT calculations~\citep{Stimpfl2006}, the unit cell shows four possible binding sites corresponding to Mg cations, three of which are closer to the surface and easily accessible to water molecules. The highest binding sites of about $19\,240$~K ($160$~kJ mol$^{-1}$) represent $45$\% of the total number of sites, while binding sites with energies around $15\,640$ and $8420$~K ($130$ and $70$~kJ mol$^{-1}$, respectively) represent $15$ and $30$\% of the total number of sites.
We created a step-like function to reproduce the surface energy distribution of a [100] crystal lattice by using three Maxwell-Boltzmann distributions with central energies at $8420$~K, $15\,640$~K and $19\,240$~K, listed in Table~\ref{tab:MCparameter}.

Each water molecule is sent randomly onto the surface and its track is recorded from the moment of its adsorption, through surface diffusion up to its eventual desorption. In this work we focus on the formation of the first monolayer, preliminary stage for water diffusion into the bulk~\citep[][submitted]{Thi2018}. The adsorption of water molecules from the gas phase occurs at a rate
 \begin{equation}
R_{\rm ads}= n_{\rm H_2O}$ $ \nu_{\rm H_2O} $ $ \sigma $ $ S  $ $ $ $ $ $ $ $ $ $ $ $ $ $ $ $ $ $ $ $ $ $\rm s^{-1},
\label{equ:accretion}
\end{equation}
which depends on the density of water molecules $n_{\rm H_2O}$, their thermal velocity  $\nu_{\rm H_{2}O}\!\sim\!0.43\sqrt{{T_{\rm gas}}/{100}}$~km s$^{-1}$, the cross section $\sigma$ of the dust surface, which scales with the size of the grid (here $6.76\,\times\,N^2_{\rm sites}\,$\AA$^2$) and the sticking coefficient $S$ (assumed equal to $1$).

If a water molecule lands in a site surrounded by neighboring H$_2$O, its binding energy increases linearly with the number of neighboring molecules~\citep{Cuppen2007} as $0.22$~eV per hydrogen bond~\citep{Dartois2013} until a maximum of $0.88$~eV is reached when a water molecule is surrounded by four neighbors. By increasing the binding energies of water molecules as they are surrounded by other water molecules, we are including the formation of water clusters in our model, and can address the contribution of such clusters for the formation of the first water layer.
\begin{table}
        \centering
         \caption[]{List of the input parameters for the Monte Carlo simulation}
      \label{tab:MCparameter}          
    \begin{tabular}{l r r}
            \hline\hline         
            \noalign{\smallskip}
            Parameter & This work & Comparison work \\
            \noalign{\smallskip}
            \hline
            \noalign{\smallskip}
            $n_{\rm H_2O}$ (cm$^{-3}$) &$6 (8)$, $6 (10)$, $6 (12)$& $1 (11) $\textsuperscript{a}\\ 
            $T$ (K)                        		& $300 - 800$ 			  & $700 - 1200$ \textsuperscript{a}\\
              				    		& 			 		  & $1000 - 1500$  \textsuperscript{e}\\
 \noalign{\smallskip}
            \hline
            \noalign{\smallskip}
      	$E_{\rm ads}$ (K) 	& $\,\,\,\,8420$ ($55$\%)   &\\
      	(100)   			& $19\,240$ ($15$\%)  	& $19\,240.4$ \textsuperscript{c}, $19\,240$ \textsuperscript{b}\\
	  				& $15\,640$ ($30$\%)   	&  $16\,120$ \textsuperscript{a}, $15\,132.4$ \textsuperscript{d}  \\ 
					               $N_{\rm sites}$  & $400$ & $32$ \textsuperscript{a}\\
        $N_{\rm grid\,cells}$ & $400$ 			        & $3924$ \textsuperscript{a},$10\,000$ \textsuperscript{e}\\
         cell size (nm$^{2}$) & $0.06700$		        & $0.00062$ \textsuperscript{a}\\
                 \hline
                 \noalign{\smallskip}
         \end{tabular}
\tablebib{
(a)~\citet{Muralidharan2008}; (b)~\citet{Stimpfl2006}; (c)~\citet{King2010}; (d)~\citet{Prigiobbe2013};
(e)~\citet{Stimpfl2004}.
}
\tablefoot{The (..) notation indicates the power of $10$.} 
 \end{table}
 
We also include surface diffusion of water molecules. Once landed on the grid, the water molecules can also move from one site to another with a diffusion rate
\begin{equation*}
R_{\rm diff}= \nu$ $ $exp$\left[-0.4 \times \displaystyle\frac{n_{\rm nb} E_{\rm b}}{T}\right] $ $ $ $ $ $ $ $ $ $ $ $ $ $ $ $ $ $ $ $ $ $\rm s^{-1},
\label{equ:diffusion}
\end{equation*}
where $\nu$ is the vibrational frequency of a water molecule in its site, that is $10^{12}$~s$^{-1}$, $E_{\rm b}$ is the energy of a single hydrogen bond, and~$n_{\rm nb}$ the number of neighbors. The activation energy for diffusion is $40$\% of the binding energy, as in~\citet{Cazaux2015}; hence, it depends on the binding site and number of neighboring water molecules.

Once adsorbed on the surface, water molecules can sublimate back into the gas phase. The desorption rate depends on the binding energy of the water molecules and is therefore directly dependent on the number of neighbors~$n_{\rm nb}$. The desorption rate of one water molecule with~$n_{\rm nb}$ neighbors can therefore be written as
\begin{equation*}
R_{\rm des} = \nu$ $$exp$\left(-\displaystyle\frac{n_{\rm nb} E_{\rm b}}{T}\right) $ $ $ $ $ $ $ $ $ $ $ $ $ $ $ $ $ $ $ $ $ $\rm s^{-1}.
\label{equ:evaporation}
\end{equation*}

While desorption rates increase exponentially with the surface temperature, accretion rates increase linearly with the density of water molecules.  
The coverage of water molecules on dust surfaces is governed by these two competing mechanisms. In the next section, we address the kinetics of forsterite hydration.  
\begin{figure}[htp]
\centering
{\includegraphics [width=\columnwidth]{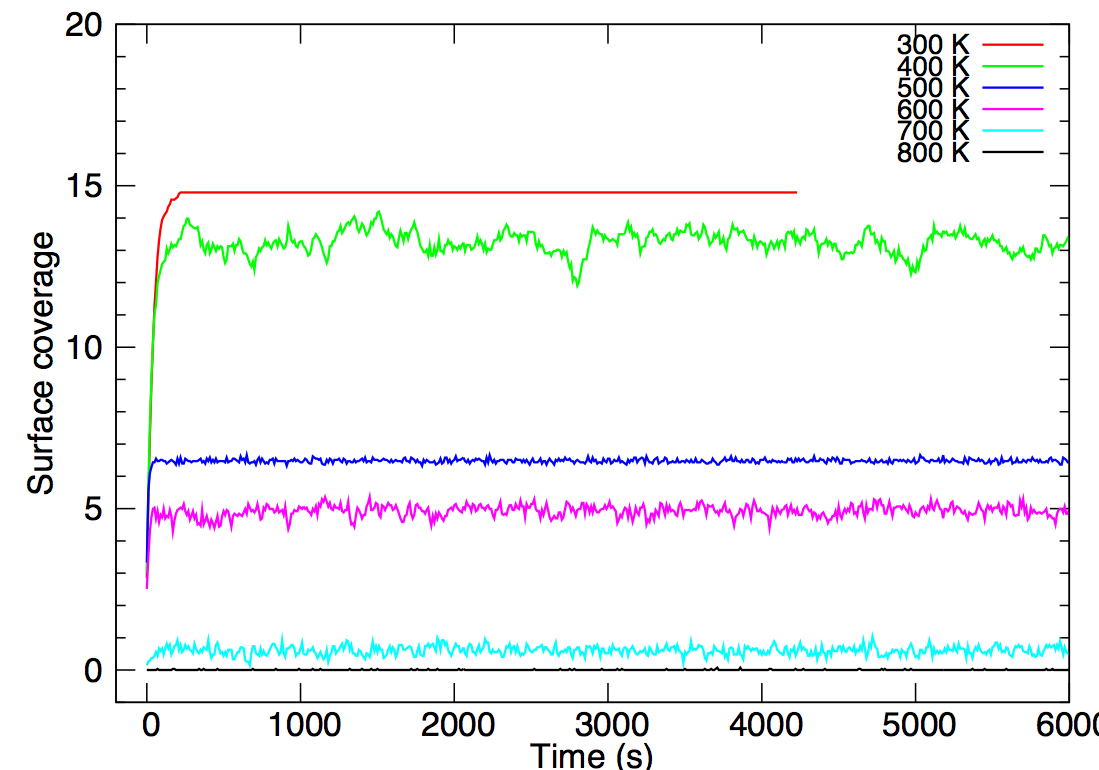}}\\
{\includegraphics [width=\columnwidth]{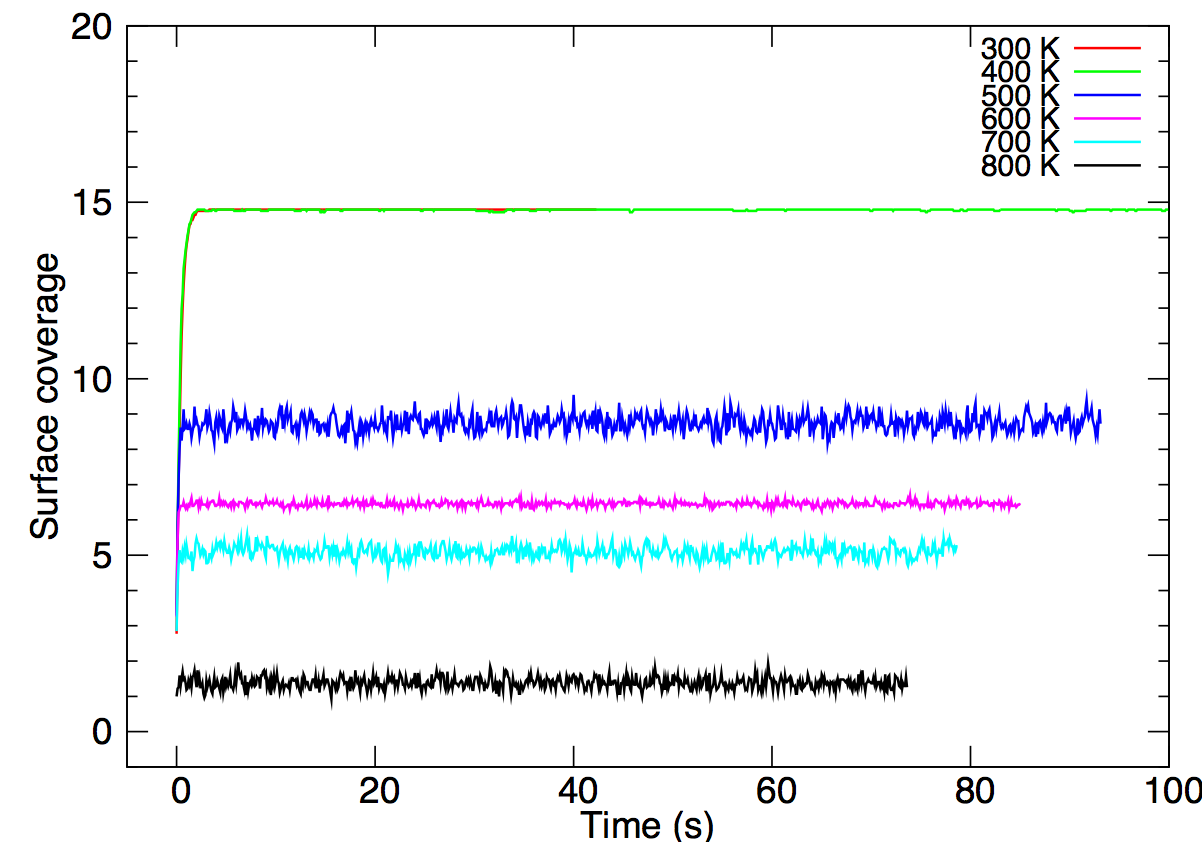}}\\
{\includegraphics [width=\columnwidth]{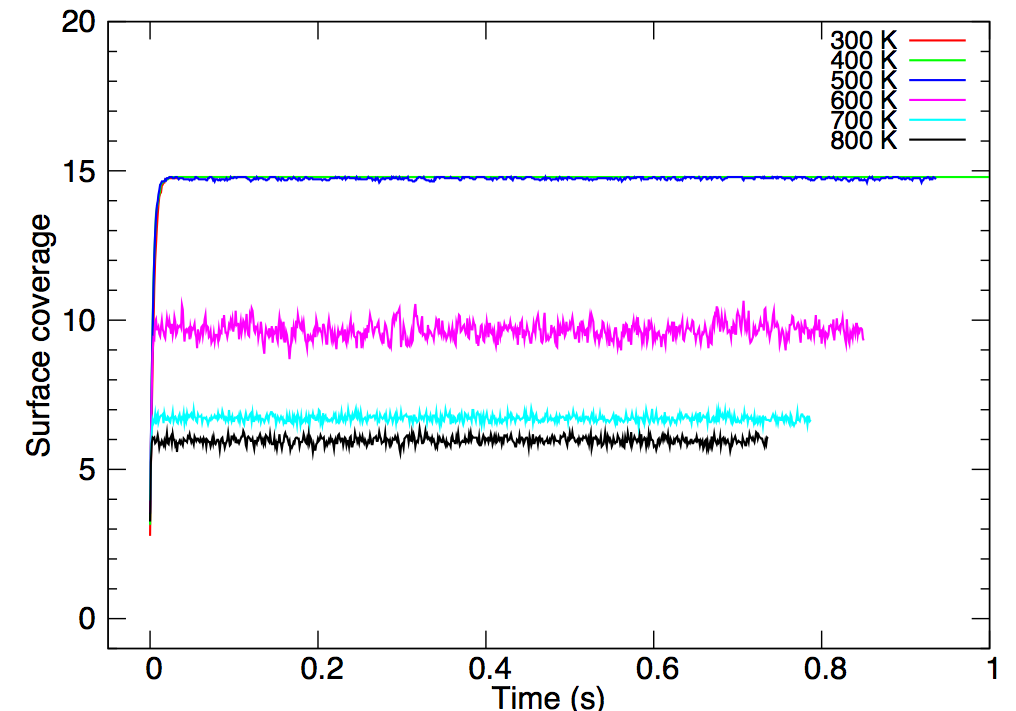}}\\
\caption{{Surface coverage (H$_{2}$O~nm$^{-2}$) as a function of time (s) at three water vapor densities: $6\,\times\,10^8$ (top), $6\,\times\,10^{10}$ (middle) and $6\,\times\,10^{12}$ (bottom)~cm$^{-3}$ and temperatures between [$300 - 800$]~K (see the legend). Except for $300$~K, at each temperature the adsorption rate increases with the water vapor density and equilibrium is reached within fractions of a second for a density of $6\,\times\,10^{12}$~cm$^{-3}$.}}
\label{fig:rates}
\end{figure}
\begin{figure*}[htbp]
{\includegraphics [width=0.93\columnwidth]{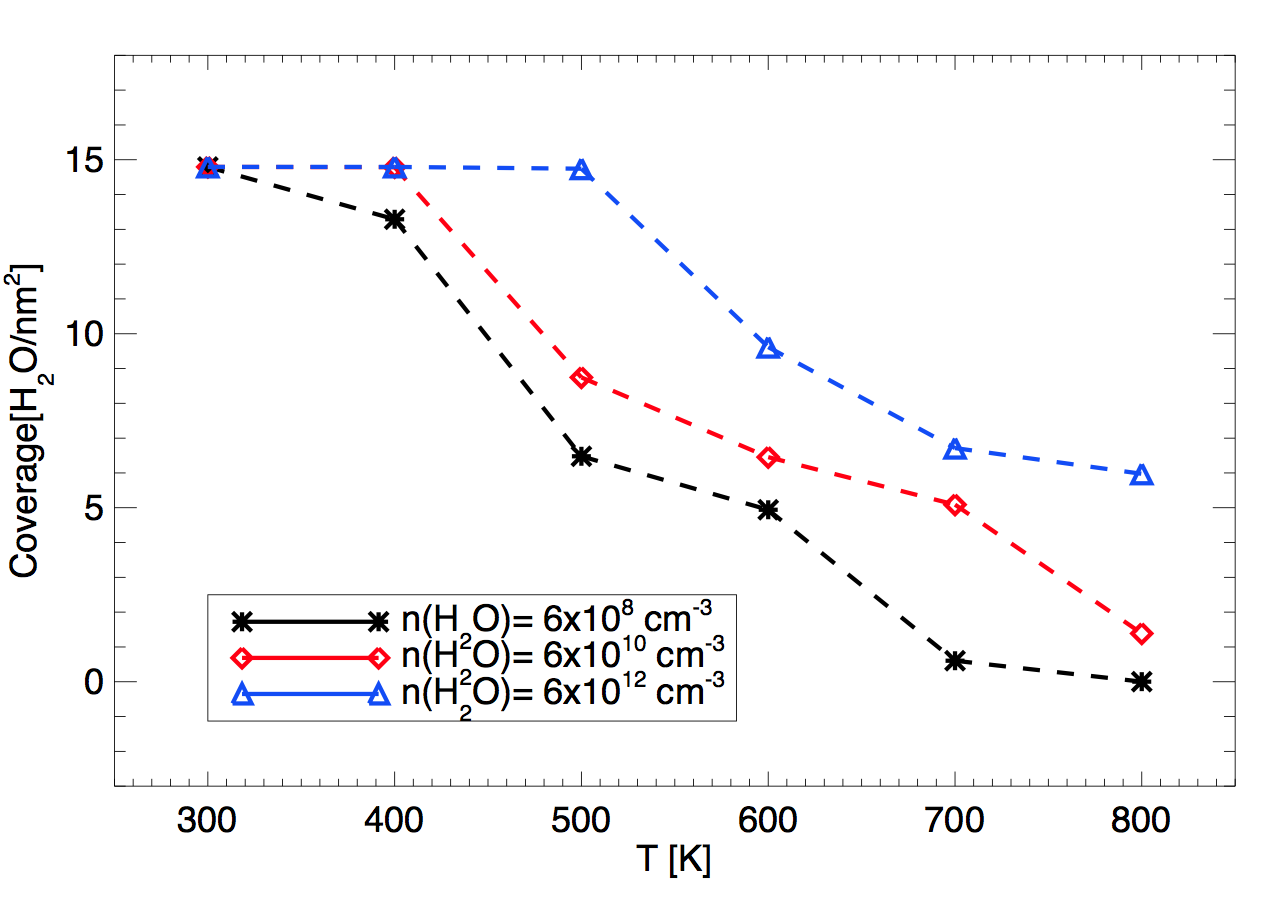}}
{\includegraphics [width=\columnwidth]{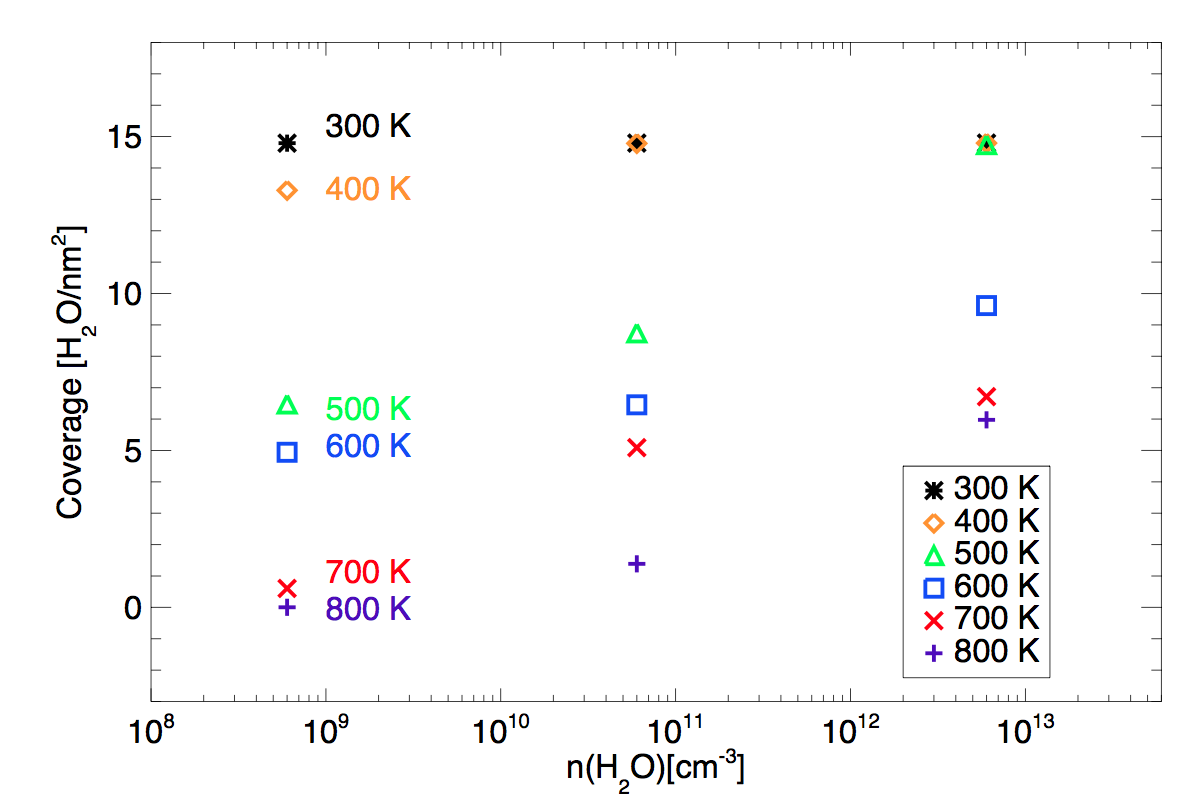}}\\
{\includegraphics [width=\textwidth]{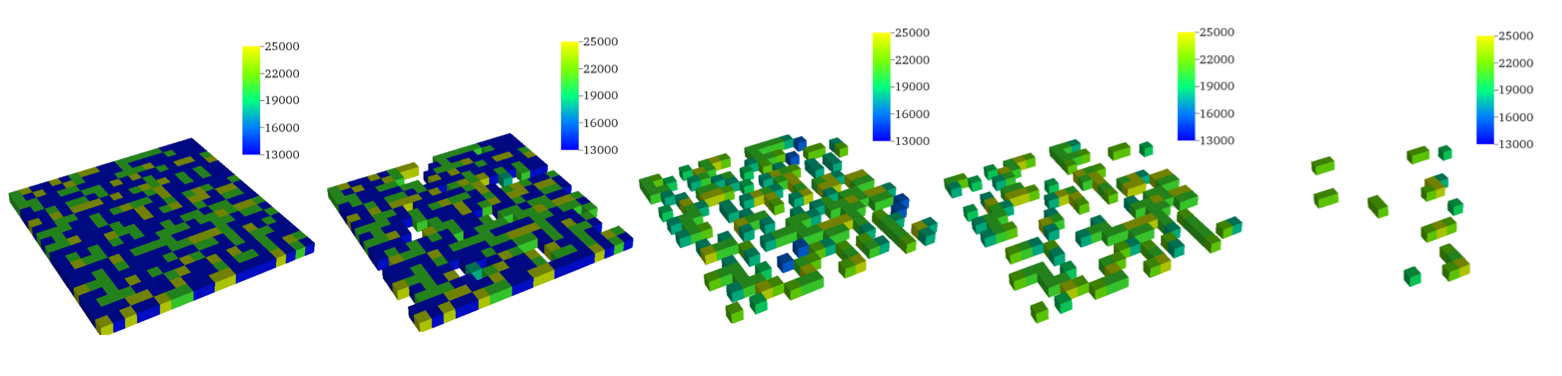}}
\caption{{Top: surface coverage (H$_{2}$O~nm$^{-2}$) as a function of temperature (K) (left) and water vapor density (cm$^{-3}$) (right). Each data point is the average $\pm\sigma$ of the coverage values at equilibrium (see Fig.~\ref{fig:rates}). Left: dashed lines are used here to link each point and emphasize the step-like decrease of the surface coverage with the temperature. Bottom: simulated [100] forsterite surface grids at $6\,\times\,10^{8}$~cm$^{-3}$ and $T\,=\,[300 - 700]$~K (from left to right). The color scale indicates the adsorption energy of the occupied sites in the range between $13\,000$~K  (blue) and $25\,000$~K (yellow).}}
\label{fig:coverage}
\end{figure*}

\subsection{Results from Monte Carlo simulation}\label{MCresults}
We have performed several simulations with equal grid properties at temperatures and water vapor densities indicated in Table~\ref{tab:MCparameter}. Our motivation is to study how much water can be accreted onto dust grains in the parameters space ($T, n_{\rm H_2O}$) that defines the region between $0.07-0.3$~au from the protosun where the raw material for terrestrial planets may have experienced morphological (thermal and aqueous) alterations.

In Figure~\ref{fig:rates}, adsorption rates of water vapor molecules onto the forsterite [100] crystal lattice are shown for various surface temperatures ( $T_{\rm gas}\!=\!T_{\rm dust}$) and water vapor densities of $6\,\times\,10^{8}$~cm$^{-3}$~(top), $6\,\times\,10^{10}$~cm$^{-3}$~(middle) and $6\,\times\,10^{12}$~cm$^{-3}$~(bottom). For the lowest water vapor density considered, at $300$~K the [100] crystal surface is fully covered by one monolayer which represents~$15$~H$_2$O~nm$^{-2}$, at $400$~K about $90$\% of the surface is hydrated; at a temperature between $500$~K and $600$~K the surface coverage drops to about $30$\% of the total surface area and at $800$~K no water molecules stick to the surface. Upon exposure to the intermediate water density (see Fig.~\ref{fig:rates} middle plot) the coverage at equilibrium increases for each considered temperature. At the highest water vapor density (see Fig.~\ref{fig:rates} bottom plot) at $800$~K one third of the forsterite grain can retain water on its surface.
The equilibrium for the formation of one monolayer is rapidly achieved within $100$~s at the lowest vapor density, within $10$~s for the intermediate density value and less than $0.1$~s for the highest vapor density considered. 

The average surface coverage at equilibrium was calculated and plotted as a function of temperature (Fig.~\ref{fig:coverage} left panel) and density (Fig.~\ref{fig:coverage} right panel). This figure illustrates the competition between accretion and evaporation, and the formation of clusters. While at low temperatures the surface is fully covered by water molecules, this coverage decreases at different paces depending on the accretion rates (densities). The higher the density, the higher the accretion rates and the ability to form water clusters (and therefore increasing the binding energies). 

The decrease of surface coverage for increasing temperatures is not linear, but reflects the step-like function used to describe the three different types of binding sites with different binding energies (see Table~\ref{tab:MCparameter}). This distribution of binding sites and the cluster effect also cause the surface coverage to decrease to zero smoothly when approaching $700$~K and $800$~K.
For an increasing $n_{\rm H_2O}$, the surface coverage increases also not linearly, as shown in Fig.~\ref{fig:coverage} right panel. This is more visible for higher temperatures and reflects the competition between desorption and adsorption of water molecules which are allowed to cluster around an occupied site.  

In our MC simulations two important surface processes were implemented: surface diffusion and water cluster formation. The latter helps water molecules to increase their binding energies as other adsorbed molecules, present in neighboring sites, will add energies to the binding through H-bonds. The effect can be seen in the lower panel of Fig.~\ref{fig:coverage}, where at equal water vapor density, molecules resist thermal desorption by forming a "cluster" around an occupied site thus increasing their adsorption energy.
As claimed in~\citet{deLeeuw2000} and~\citet{Stimpfl2006}, this cooperative behavior increases the chances for more water to adsorb and to remain on the surface, in particular at high temperatures.

\subsection{Discussion of MC simulations}\label{MCdiscussion}
In case of a gas rich in water vapor, molecular adsorption onto silicate grains would be an efficient hydration mechanism over a wide temperature range within planetary accretion time-scales.

Our results deviate somewhat from previous works. For instance, at $700$~K and $6\,\times\,10^{10}$~{H}$_2$O~cm$^{-3}$, similar physical parameters used by~\citet{Muralidharan2008}, our simulation shows that only $30$\% of the [100] forsterite surface is occupied by water molecules, while~\citet{Muralidharan2008} obtained full surface coverage. Our equilibrium time scales are three orders of magnitude smaller than the one obtained in~\citet{Muralidharan2008} simulations (within $20\,000$~s at all temperatures in the range [$700 - 1200$]~K), which is very short compared to the nebula lifetime of millions of years.

Finally, in our work the surface coverage decreases in steps with the temperature, in contrast with the results of~\citet{Muralidharan2008}, where an exponential trend was found for the temperature range $[700-1200]$~K.
These discrepancies can be due to differences between the two models and input parameters.
Adsorption and a "box" of gas molecules in the random-walk regime with equal mean free path and collision frequency to the [100] forsterite surface are treated in the same way in this work and in~\citet{Muralidharan2008}. 
In~\citet{Muralidharan2008}, the surface potential is detailed finely using a grid sub-sampled in $3924$ cells of area $0.062$~\AA$^2$ compared to the $400$ sites of $6.76$~\AA$^2$ for our model surface. This can explain the smooth, exponential decrease of the surface coverage with the temperature versus the step-like, non-linear trend in our simulations. However, our temperature range overlaps with theirs only for two values. Besides, we allow the water molecules to scan the surface upon collision and, by lateral diffusion, to find the favorable binding site. This can explain why the time spent for the system to reach equilibrium is a factor of $1000$ shorter than in~\citet{Muralidharan2008}, where a simple collision theory is considered. 

To further understand the reason for the differences, we have investigated different surface potential energy distributions (Appendix~\ref{appendixII}), reproducing the energy potential in~\citet{Stimpfl2006} and used in~\citet{Muralidharan2008}. Again, at $700$~K and $6\,\times\,10^{10}$~{H}$_2$O~cm$^{-3}$ no full coverage was attained contrary to the results from~\citet{Muralidharan2008}. The cause of this discrepancy can depend on an oversampling of the surface sites which can affect the energy distribution. 

\section{Solar nebula implications}\label{Astroimplications}
Several near- and mid-IR observations~\citep{Salyk2008,Carr2008,Pontoppidan2010,Carr2011,Riviere2012} have revealed the presence of warm water vapor ($300 - 800$~K) in the habitable planet-forming region (within three astronomical units) in T Tauri circumstellar disks.  
Embedded in this environment, small micron-sized silicate dust grains have also been observed from their signatures at $10~\mu$m and $18~\mu$m~\citep[e.g.][]{Kessler-Silacci2006}. Shape, intensity and exact wavelength of these spectral features are indicative of the dust morphology (crystalline or amorphous), mineralogy (i.e. forsterite, enstatite, etc.) and size properties~\citep[see for example][]{Jager2003, Chiang2004, Bouwman2008}.

Despite the long lasting debate on which mechanisms contributed to the water content on rocky planets of our Solar System (exogenous vs. endogenous), our work shows that water gas-solid interaction can lead to hydration of the pristine forsterite surface under nebula conditions within its lifetime. In the following, the surface coverages obtained from our MC simulations are discussed in the context of the Earth mantle water and oceans, and compared to the amount of hydrated silicates found in meteorites and observed on asteroids.

It is clear that our approach has its limitations as we neglect several processes that affect both the global and local grain size distribution in the nebula. Grain growth (and destruction), vertical settling and radial migration are processes that act in young disks as they are evolving into planetary systems, see e.g. \citet{Birnstiel2016} for a recent review. \citet{Pinte2016} show that ALMA images of the young T Tauri disk around HL Tau indicate the presence of large mm-sized grains that have settled efficiently to the midplane. On the other hand, \citet{Kruijer2014} find that the parent bodies of various types of iron meteorites likely accreted within 0.1-0.3~Myr after CAIs, thus providing support to the possibility that large bodies could already exist also in protoplanetary disks with ages of 1-3~Myr. In that context, the assumptions made below of either single small grain size or an observed homogeneous grain size distribution throughout the inner disk can only be limiting cases and more detailed studies are required in the future to combine the results from water adsorption with detailed dust evolutionary models of the inner disk.

\begin{figure}[htp]
\centering
{\includegraphics [width=\columnwidth]{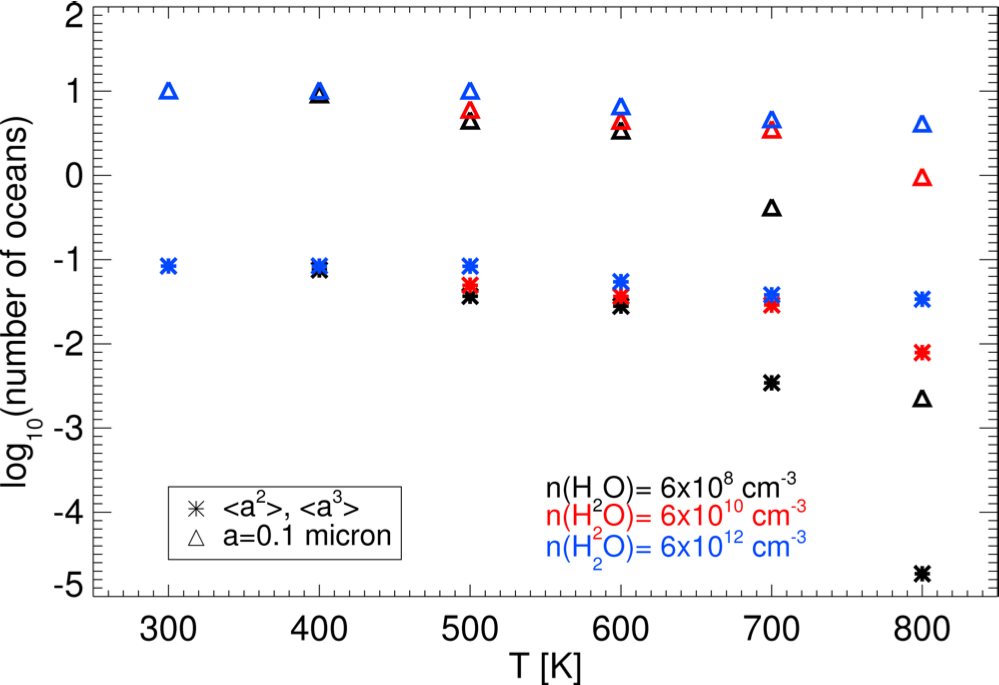}}
\caption{{The conversion of the surface coverages (Fig.~\ref{fig:coverage}) into number of Earth's oceans as a function of the temperature in the range [$300-800$]~K and for the three water density values~$6\,\times\,10^{8}$~cm$^{-3}$~(black),~$6\,\times\,10^{10}$~cm$^{-3}$~(red) and~$6\,\times\,10^{12}$~cm$^{-3}$~(blue). The calculation is done considering spherical dust grains of $0.1~\mu$m average radius ($\triangle$) and the second, $\langle a^2 \rangle$, and third moments, $\langle a^3 \rangle$, ($\textasteriskcentered$) derived from the grain size distribution.}}
\label{fig:ocean}
\end{figure}
\subsection{Water content on Earth}\label{oceans}
The oceans are not the only water reservoir of our planet. Adding together the water contributions of the hydrosphere ($1.6\,\times\,10^{21}$~kg), the exosphere including the crust ($1.9\,\times\,10^{21}$~kg) and the mantle (between $0.3-2.5\,\times\,10^{21}$~kg) and normalizing by the mass of the Bulk Silicate Earth (BSE = mantle + crust,~$4.05\,\times\,10^{24}$~kg), the total water budget ranges between $0.05-0.11$~wt\%. The Earth is currently~$5-50$ times dryer than the CV and CO, the two driest classes of C-chondrites~\citep{Mottl2007}. Nevertheless, the $wet$ bodies that collided to form our protoplanet must have possessed a larger amount of water than the actual Earth's abundance to account for the loss of such volatile species by impact degassing.

We have estimated the Ocean Equivalent Water (OEW) by combining our knowledge of the Earth's radius ($R_\oplus\!=\!6.371\,\times\,10^6$~m) and the total oceans mass ($1.4\times10^{24}$~g) together with the surface coverage data, here called~$\theta$~({H}$_2$O~nm$^{-2}$), obtained from our MC models (see Sec~\ref{MCresults}). The water mass (g) adsorbed onto a spherical grain with radius $a$ and surface area $4\pi a^2$ is
 \begin{equation}
\rm M_{\rm H_2O} = 4\pi$ $a^2 $$m_{\rm H_2O}$ $ \theta ,
\label{equ:masswater}
\end{equation}
where  $m_{\rm H_2O}$ is the water molecular mass.
The number of grains equivalent to the Earth's volume is
 \begin{equation}
N_{gr} = \frac{R_\oplus^3}{a^3}.
\label{equ:numbergrains}
\end{equation}
Multiplying $N_{gr}$ by $\rm M_{\rm H_2O}$ and normalizing for the oceans mass, we obtain the number of oceans that could originate from the agglomeration of $N_{gr}$ forsterite dust grains previously hydrated by water vapor condensation. 

In Figure~\ref{fig:ocean}, the results are plotted as a function of temperature in the range [$300-800$]~K, for the three water vapor densities studied in this work, $6\,\times\,10^{8}$~cm$^{-3}$~(black), $6\,\times\,10^{10}$~cm$^{-3}$~(red) and $6\,\times\,10^{12}$~cm$^{-3}$~(blue). Two sets of calculations are shown: One where we assume that the dust grains are all spheres of $0.1~\mu$m radius (see $\triangle$ symbol in Fig.~\ref{fig:ocean}). A second one, where hydration occurs on grains which have previously agglomerated into larger bodies according to the power-law size distribution described in Section~\ref{prodimoinputs}.
These two scenarii present the extremes of assuming that grains can retain their water during the agglomeration process and grains only starting to take up water after the first agglomeration phase.

In the first scenario (see $\triangle$ symbol in Fig.~\ref{fig:ocean}), about $10$~Earth oceans could be delivered by nebular adsorption of water molecules at $300$~K and $400$~K for the three densities considered.
For the lowest density value (black triangle in Fig.~\ref{fig:ocean}), $4.5$, $3.5$, 0.5 and no Earth oceans can form from water vapor condensation respectively at $500$~K, $600$~K, $700$~K and $800$~K. The number of oceans scales linearly with the surface coverage $\theta$ (see Eq.~\ref{equ:masswater}), which increases with the density (see Fig.~\ref{fig:coverage} and Sec.~\ref{MCresults}). Therefore, the Earth can still inherit between one and four oceans from the agglomeration of wet $0.1~\mu$m sized grains upon exposure at $800$~K to water molecules at densities of~$6\,\times\,10^{10}$~cm$^{-3}$~(red triangle) and~$6\,\times\,10^{12}$~cm$^{-3}$~(blue triangle), respectively.

Assumed that the [100] surface constitutes about $44$\% of the surface area of a perfect forsterite crystal, for a water density of $10^{11}$~cm$^{-3}$~\citet{Muralidharan2008} obtained between eight and four Earth oceans at respectively $700$~K and $800$~K, which are a factor of five to nine larger than our results.
In a later work,~\citet{Asaduzzaman2015} showed the trend of the OEW as a function of grain size for different coverage and at $900$~K and $10^3$~bar of water pressure (likely Earth forming condition).

Considering that $65$\% of the Earth's mantle is olivine mineral, agglomeration of $0.1~\mu$m grains would account for about five and two OEW in the Earth's mantle, respectively for full or partial coverage ($42$\% of the total surface area). This occurs with most of the ($T, n_{\rm H_2O}$) pairs we considered in our models.

In~\citet{Stimpfl2004}, Monte Carlo simulations are used to evaluate the coverage of water on a substrate on a grid of $10\,000$~cells. These authors find that the adsorbed water potentially stored in the dust corresponds to about three times the Earth's oceanic + atmospheric + crustal water (OAC) and about $1.5$ times the Earth's OAC + mantle water. According to the latest Earth's water estimation reported by~\citet{Genda2016}, roughly four ocean masses are needed to account for the "surface" and mantle water. Our results demonstrate that during the early solar nebula at a distance of $0.07-0.3$~au from the protosun and with the parameters space ($T, n_{\rm H_2O}$) described earlier, $0.1~\mu$m dust grains are subject to an intense aqueous alteration.
By the same mechanisms that transport processed materials in- and outwards in the disk~\citep{Gail2004, Boss2004, Nuth2005}, these wet silicates would eventually reach the terrestrial planets feeding zone (within one astronomical unit) and there supply enough water for oceans to rise.

In our ProDiMo models, the grain size distribution holds for the entire disk with the size ranging between $0.05$ and $3000$~$\mu$m (see Table~\ref{parameter}). However, dust settling vertically to the midplane changes the second and third moments of the distribution. 
Accordingly, in Equations~\ref{equ:masswater} and~\ref{equ:numbergrains} we chose to use respectively the second and third moments of the grain size distribution (see Sec.~\ref{prodimoinputs}), which are $\langle a^2 \rangle\!=\!1.245\,\times\,10^{-10}$~cm$^2$ and $\langle a^3 \rangle\!=\!1.525\,\times\,10^{-13}$~cm$^3$. The number of Earth oceans thus obtained are two order of magnitudes smaller than in the previous case (see $\textasteriskcentered$ symbol in Fig.~\ref{fig:ocean}). In particular, between $3.5$\% to $8$\% of an ocean of water can accrete at $500$~K and between $0.5$\% and $4$\% at  $700$~K. These results are close to the ones of~\citet{Stimpfl2004}, where $1$\% and $3$\% of one Earth's ocean could accrete at $700$~K and $500$~K respectively~\citep[see][]{Drake2005}.

Hence, the contribution to the water content on habitable planets provided by hydrated silicates varies with the size distribution of the mineral grains: At a very early stage of our solar nebula, when grains where ISM-like, Earth could potentially inherit an amount of water equal to $10$ oceans. At a late stage in the disk evolution, when dust grains grow, water vapor condensation can contribute to less than $1$\% of an Earth ocean. This is a lower limit of the amount of water that can be incorporated into the grain due to the assumptions of this work. Only a single layer of water molecules was allowed to form onto a defect-free crystalline silicate surface. It was ignored that hydration is enhanced when defects are present and amorphous silicates are considered instead~\citep[][]{Yamamoto2016}. Finally, we have neglected bulk diffusion into the core of the silicate grain, which has been recently shown to efficiently occur at temperatures as high as $700$~K in the inner regions of a protoplanetary disk~\citep[][submitted]{Thi2018}. All these factors can significantly enhance the amount of phyllosilicates that can form.

\subsection{Phyllosilicates in asteroids}\label{asteroids}
For a simple estimate of the fraction of phyllosilicates in asteroids under the scenario of hydration by water vapor, we assess which fraction of the precursor grains could be turned into phyllosilicates through surface reactions. The second moment of the grain size distribution $\langle a^2 \rangle\!=\!1.245\,\times\,10^{-10}$~cm$^2$ provides the average surface of a grain. With a density of surface sites of $N_{\rm sites}$\!=\!$1.5\,\times\,10^{15}$~cm$^{-2}$, we find a total of $2.35\,\times\,10^6$ surface sites. The dust to gas mass ratio is 
\begin{equation}
\frac{M_{\rm d}}{M_{\rm H}}=\frac{n_d \cdot 4 \pi\,\langle a^3 \rangle\rho_{\rm gr}}{n_{\langle H \rangle}\,\mu\,m_{\rm H}}=0.01,\\
\end{equation}
where $M_{\rm d}$ is the mass of dust, $M_{\rm H}$ the mass of the gas (hydrogen gas),  $n_{\rm d}$ the volume density of dust, $\rho_{\rm gr}$ the mass density of dust grains, which is  2.076$~$g\,cm$^{-3}$, $\mu$ is the mean molecular weight of the gas, ($2.4$ for H$_2$ gas), n$_{\langle H \rangle}$ is the total hydrogen number density in the gas and $m_{\rm H}$ the mass of hydrogen ($1.67\,\times\,10^{-24}$ g). The volume density of dust grain can then be written as
\begin{equation}
n_{\rm d} =\frac{3 \cdot 0.01\mu\, m_{\rm H}\, n_{\langle H \rangle}}{4\,\pi\,\langle a^3 \rangle\rho_{\rm gr} }
\end{equation}
with the third moment of the grain size distribution $\langle a^3 \rangle\!=\!1.525\,\times\,10^{-13}$~cm$^3$. Knowing the volume density of dust grains, we can estimate the number of sites on the surface of the grains per cm$^{-3}$ as
\begin{eqnarray} 
n_{\rm d} \times n_{\rm sites} = \\ \nonumber
 \frac{3 \cdot 0.01 \mu\,m_{\rm H}\,n_{\langle H \rangle}}{4\,\pi\,\langle a^3 \rangle\rho_{\rm gr}} \times {4\,\pi\,\langle a^2 \rangle} \times {N_{\rm sites}}=7.08 \cdot10^{-8} n_{\langle H \rangle}
\end{eqnarray}

We assume that no diffusion of water occurs and reactions are limited to the surface. This is a conservative lower limit to the phyllosilicate production since grains will be irregularly shaped and defects at the surface will help water to diffuse into the interior. 
Assuming the stoichiometry of the following reaction
\begin{equation}
{\rm 2\,Mg_2SiO_4~+~3\,H_2O~\rightarrow~Mg(OH)_2~+~Mg_3Si_2O_5(OH)_4}\,\,\,, \nonumber
\end{equation}
and that $60$\% of the surface is silicates and $30$\% of sites are occupied by water, the lower limit to the number density of phyllosilicates on surfaces is $1.274\,\times\,10^{-8}\,\rm n_{\langle H \rangle}$~cm$^{-3}$. The fraction of water contained in dust can be written as:
\begin{eqnarray}
\frac{\rho_{\rm water\,on\,grain}}{\rho_{\rm dust}}=\\ \nonumber
\frac{3\cdot0.01\,\mu\,m_{\rm H}\,n_{\langle H \rangle}\cdot N_{\rm sites}\cdot4\,\pi\,\langle a^2 \rangle \cdot0.6\cdot0.3\cdot18\,m_{\rm H}}{4\,\pi\,\langle a^3 \rangle\,\rho_{\rm gr}\cdot0.01\,\mu\,m_{\rm H}\,n_{\langle H \rangle}}
\end{eqnarray}
This translates into a lower limit to the fraction of adsorbed water of $\sim 10^{-5}$. 
If the grains had been all $0.1~\mu$m in size, this number would be $\sim 10^{-3}$, two orders of magnitude higher.

Our models show that hydration in meteorites parent bodies (the asteroids) could have occurred in the inner and warm solar nebula. Among different types, CM and CI carbonaceous chondrites typically contain $5 - 15$\% H$_{2}$O/OH by weight~\citep[][and references therein]{Rivkin2002} with the least hydrated CVs type showing abundance of hydrogen typically below $0.5$ wt\%~\citep[][and references therein]{Beck2014}. To account for such large degree of aqueous alteration, subsequent hydration mechanisms need to be considered. For instance, diffusion of water molecules into the silicate bulk enables a higher formation rate of phyllosilicates~\citep[][submitted]{Thi2018}, in particular in those relatively hot regions (T > $700$~K) of the disk midplane where the cluster effect explored in our models does not retain water on the surface efficiently. 

\section{Conclusions}

In this work we have investigated the efficiency of water vapor adsorption onto forsterite grains surfaces as one of the mechanisms that contributed to the water on Earth and in asteroids. 

The astrophysical disk model ProDiMo tailored to the solar nebula properties was combined with Monte Carlo simulations of water adsorption on a [100] forsterite crystal lattice. Water vapor abundances, temperature and pressure radial profiles identify the region in the warm disk midplane, between $0.07 - 0.3$~au from the protosun, where hydration of dust grains could have occurred. Several MC simulations were run to assess the dependency of the adsorption rate and the surface coverage on the parameter space identified by the pairs ($T, n_{\rm H_2O}$).

Our MC models show that complete surface water coverage is reached for temperatures between $300$ and $500$~K. For hotter environments ($600$, $700$ and $800$~K), less than $30$\% of the surface is hydrated.
At low water vapor density and high temperature, water cluster formation plays a crucial role in enhancing the coverage (see also Appendix~\ref{appendixIII}). The binding energy of adsorbed water molecules increases with the number of occupied neighboring sites, enabling a more temperature-stable water layer to form. Lateral diffusion of water molecules lowers the timescale for surface hydration by water vapor condensation by three order of magnitude with respect to an SCT model, ruling out any doubts on the efficiency of such process in a nebular setting.

Finally, the amount of water potentially delivered on Earth drastically varies if we rely on a grain size distribution instead of single sized grains. Grain agglomeration and dust settling to the midplane, the initial steps for planetesimal formation, should clearly lead to a wide grain size distribution as the nebula evolves. In order to improve our initial estimates, detailed dust evolution models should be combined with the water adsorption efficiencies found here.

In addition, dynamical simulations of grain growth are required to understand how agglomeration and collision processes affect the amount of water retained on the grain surfaces and how this competes with the diffusion timescale of water molecules into the bulk of the grains. 

\begin{acknowledgements}

This work is part of the Dutch Astrochemistry program financed by the Netherlands Organisation for Scientific Research, \emph{NWO}. 

\end{acknowledgements}
\bibliographystyle{aa}  
\bibliography{mybib} 
\Online
\begin{appendix}
\section{H$_2$ as gas competitor}\label{appendixI}

In the following, we will make some simple estimates to assess the role of dust and H$_2$ in our Monte Carlo simulations. The simplified approach assumes that we have a constant influx of water molecules onto a fixed surface (representing part of the surface of a single grain). However, the water molecules encounter within a fixed volume other dust grains and H$_2$ molecules. In a primordial disk, water is typically four orders of magnitude less abundant than H$_2$. 
%
This implies that water will repetitively collide with H$_2$ before reaching a dust grain, which could change the time scale for water molecules to reach the dust. 
As H$_2$ collides frequently with the grain surface it could hinder or block the influx of water. In this study, we consider that H$_2$ molecules can only be physisorbed on surfaces. H$_2$ chemisorption is a dissociative process with a high barrier \citep{rice1987}, which makes the sticking probability of H$_2$ as two chemisorbed H atoms negligible. We therefore neglect H$_2$ chemisorption, and consider only the effect of the grains covered by physisorbed H$_2$ molecules and estimate if this coverage could influence the sticking of water. At the typical gas temperatures considered here, H$_2$ cannot stick to the warm silicate surface. However, even an extremely short residence time could block that surface site from adsorbing water. The following estimates relate to one representative set of conditions in the disk with $n_{\rm H_2}\!=\!10^{15}$~cm$^{-3}$ and $T_{\rm g}=500$~K. The size of an adsorption site is assumed to be $(2.6)^2$~\AA$^2$ and the thermal velocity of H$_2$ molecules is 
\begin{equation}
v_{\rm H_2}^{\rm th} = \sqrt{\frac{2kT_g}{\pi m_{\rm H_2}}} = 1.15\,10^5~{\rm cm/s}\,\,\,.
\end{equation}
The influx of H$_2$ into a surface site can then be calculated to be
\begin{equation}
F_{\rm H_2} = n_{\rm H_2} v_{\rm H_2}^{\rm th} (2.6\,10^{-8})^2 = 7.77\,10^4~{\rm s}^{-1}\,\,\,.
\end{equation}
At the same time, the influx of water molecules is $2.60$~s$^{-1}$ with $v_{\rm H_2O}^{\rm th}\!=\!3.83\,10^4$~cm/s. 
The residence time on the surface for H$_2$ molecules is given by the rate constant for adsorption
\begin{equation}
\tau = \frac{1}{\nu^{\rm osc} \exp{(-E^{\rm ads}/kT_{\rm d}})}\,\,\,
\end{equation}
with the adsorption energy $E^{\rm ads}/k\sim500$~K for H$_2$ physisorption~\citep{Pirronello1997}, the lattice vibrational frequency $\nu^{\rm osc}\!=\!(2 n_{\rm surf} k E^{\rm ads}/(\pi^2 m_{\rm H_2}))^{1/2}\!=\!2.50\,10^{12}$~s$^{-1}$, and the density of surface sites $n_{\rm surf}\!=\!1.5\,10^{15}$~cm$^{-2}$. Assuming furthermore $T_{\rm d}\!=\!T_{\rm g}$ yields $\tau\!=\!1.08\,10^{-12}$~s.
If an H$_2$ molecule would be present on the arrival site of a water molecule, then the water molecule would occupy the place of the H$_2$ and swap~\citep{Cuppen2007}. In any case, it has been proven experimentally that a surface covered by H$_2$ molecule would be more accommodating to incoming species and therefore that the sticking coefficient would be increased. Molecular hydrogen covering a surface serves as a medium to loose kinetic energy and to become more easily thermalised on the surface and adsorb~\citep{Gavilan2012}.
\section{Surface potential energy distribution}\label{appendixII}

\begin{figure*}[htbp]
\centering
{\includegraphics [width=0.65\columnwidth]{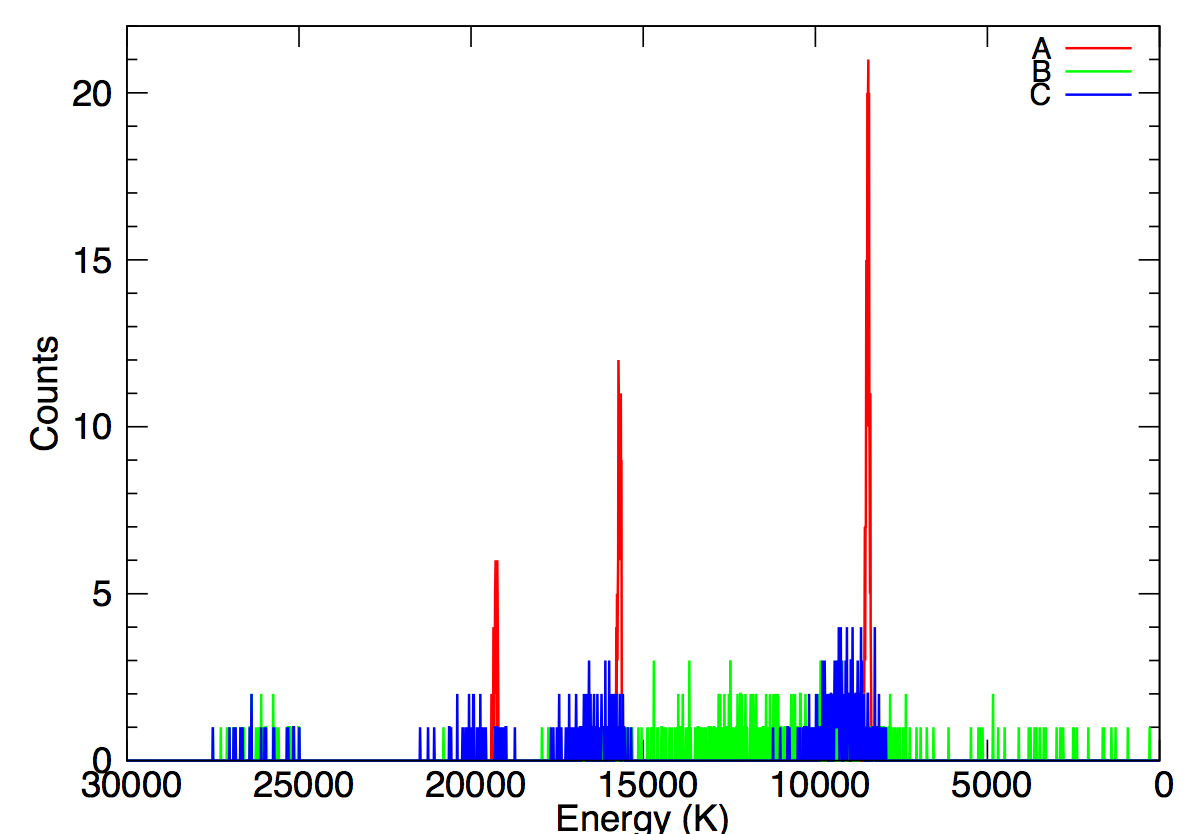}}
{\includegraphics [width=0.65\columnwidth]{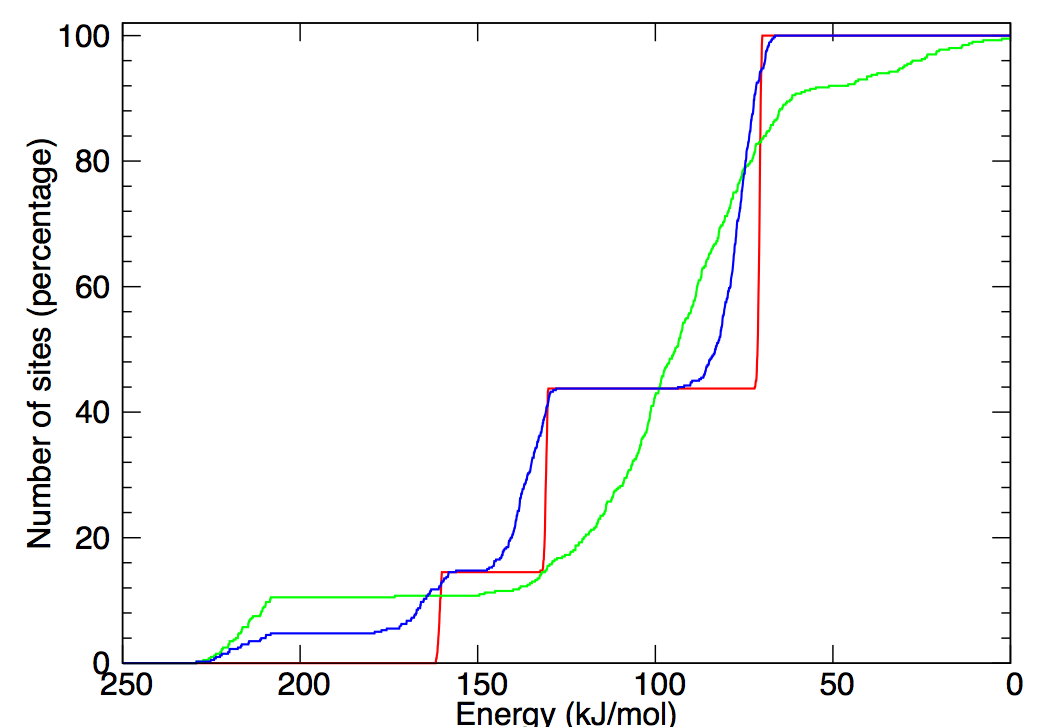}}
{\includegraphics [width=0.65\columnwidth]{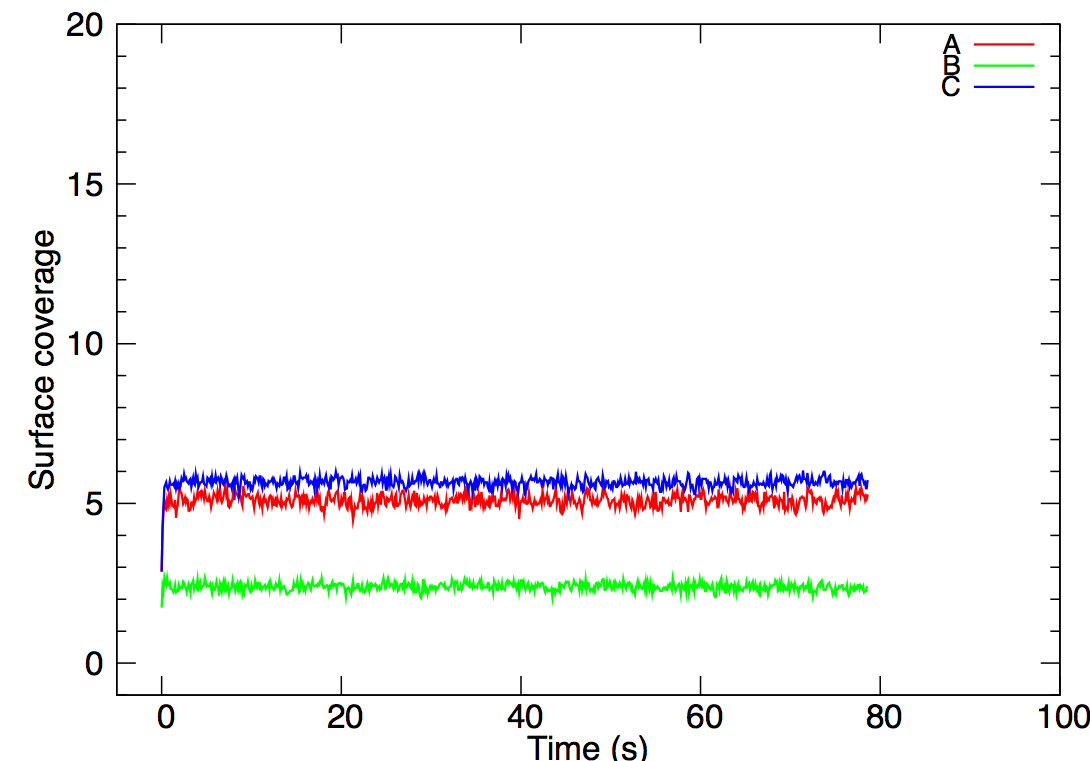}}
\caption{{Surface potential energy distribution (left plot) and surface sites density (middle plot) of three models: A (red curve), B (green curve) and C (blue curve). The surface coverage for the three models is calculated at $700$~K and $n_{\rm H_2O}\!=\!6\,\times\,10^{10}$~cm$^{-3}$ (right plot).}}
\label{fig:potential}
\end{figure*}

In the attempt to explain the discrepancies between our results and the ones obtained by~\citet{Muralidharan2008}, we have simulated the water adsorption mechanism on the forsterite [100] crystal plane using three surface energy potential models, called A, B and C (Fig.~\ref{fig:potential}). These are obtained by three Maxwell-Boltzmann distributions with parameters listed in Table~\ref{tab:MCappendix}.

Model A (red curve in the left and middle plots of Fig.~\ref{fig:potential}) is a step function composed by three distributions, where only $45$\% of the sites have binding energies greater than $14\,430$~K ($120$~kJ~mol$^{-1}$), hence those most favorable for attracting the first water molecules~\citep{Stimpfl2006}. 

Model B (green curve) is constructed by matching the surface potential energy distribution shown in~\citet{Stimpfl2006}. Here sites with binding energies smaller than $8420$~K ($70$~kJ~mol$^{-1}$) are included, contrary to model A. 

Finally, the model C (blue curve) includes $8$\% of surface sites with adsorption energies higher at $25\,260$~K ($210$~kJ~mol$^{-1}$), which were not taken into account in model A although claimed by the atomistic model of~\citet{Stimpfl2006}.
The surface energy potential distribution of the adsorption sites of the crystal surface determines the coverage of the adsorbed water molecules. As shown in the right plot of Fig.~\ref{fig:potential}, in model C the surface coverage is $5$\% higher than in model A at $700$~K and $n_{\rm H_2O}\!=\!6\,\times\,10^{10}$~cm$^{-3}$. The surface potential energy distribution of model B allows only half the coverage than seen in model A and C, due to the higher hydrophobicity of the surface.

We can conclude that the choice of the sites adsorption energy does not explain the factor of three difference in surface coverage between our work and the~\citet{Muralidharan2008} simulations.

\begin{table}
        \centering
         \caption[]{List of parameters (central adsorption energy, sigma and intensity) of the Maxwell-Boltzmann distributions of surface energy potential models called A , B and C.}
      \label{tab:MCappendix}          
    \begin{tabular}{l c r}
     \hline\hline         
            \noalign{\smallskip}
           E$_{\rm bin}$ (K) & $\sigma$ & (\%) \\
            \noalign{\smallskip}
            \hline
            \noalign{\smallskip}
            & \textbf{Model A} \textsuperscript{a}&\\
            \midrule
           $19\,240$ ($160$)& $80$ &$15$\\      
            $15\,640$ ($130$)&$80$  &$30$\\
           \,\, $8\,420$ ($70$) &$80$ & $55$\\
\noalign{\smallskip}
\midrule
           & \textbf{Model B}&\\
	\midrule
         $25\,260$ ($210$)   & $1\,000$&$10$\\      
            \,\, $8\,420$ ($70$) & $4\,000$&$80$\\
            \,\,\,\,\,\,\,\,\,\,\,\,  $0$ \,\,(\,$0$)& $4\,000$&$10$ \\
\midrule
           & \textbf{Model C}&\\
	\midrule
        $25\,260$ ($210$)    &$1\,000$&\, $8$\\      
         $16\,840$ ($125$)   & \,\, $500$&$22$\\
         \,\, $8\,420$ ($70$)   &$1\,000$ &$60$ \\  
 	 \,\, $3\,010$  ($25$) &$2\,000$ &$10$\\  
 \noalign{\smallskip}
        \hline
         \end{tabular}
\tablefoot{Values in (..) are expressed in~kJ~mol$^{-1}$.
\tablefoottext{a}{Model used in the paper.}
} 
 \end{table}
\section{Water Cluster simulation}\label{appendixIII}

In our MC models the water molecules randomly "walk" on the surface until they adsorb at a favorable site. As two water molecules sit on neighboring sites, a hydrogen bond links the two species, and a dimer can be created with the total binding energy higher than the ones of single molecules. This generates the formation of water clusters for which the binding energy increases with increasing size~\citep{Gonzales2007}. In two dimensions, the binding energies increase with the number of water molecules present in the cluster, while for a $3$D cluster~\citet{Lin2005} show that the binding energy mainly depends on the number of water molecules close to the surface. 

In this work, we used a simple way to calculate the binding energies, which increase linearly with the number of neighboring water molecules~\citep{Dartois2013}, but limiting the number of water molecules to four. 
Water clusters have been implemented in previous MC simulations to study the formation of water on carbon grains in the ISM~\citep{Cazaux2010}, to investigate the porous structure of ices~\citep{Cazaux2015} and its effect on the location of the snow line in different astronomical environments~\citep{Cazaux2011}.

Temperature programmed desorption experiments conducted by~\citet{Brown2007} confirm that the desorption of pure H$_2$O ice on a graphite surface occurs at a higher temperature for increasing coverage. 
Therefore, as the coverage increases, the clusters become more important, and a higher surface temperature is needed to desorb the molecules.

\begin{figure*}[htbp]
\centering
{\includegraphics [width=\columnwidth]{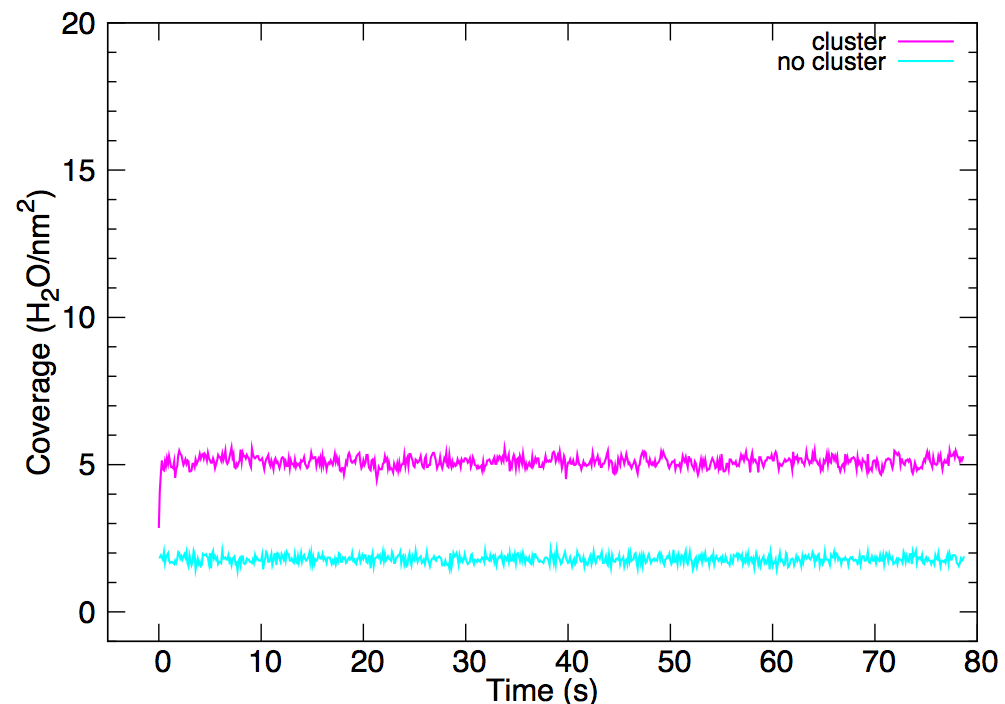}}
{\includegraphics [width=\columnwidth]{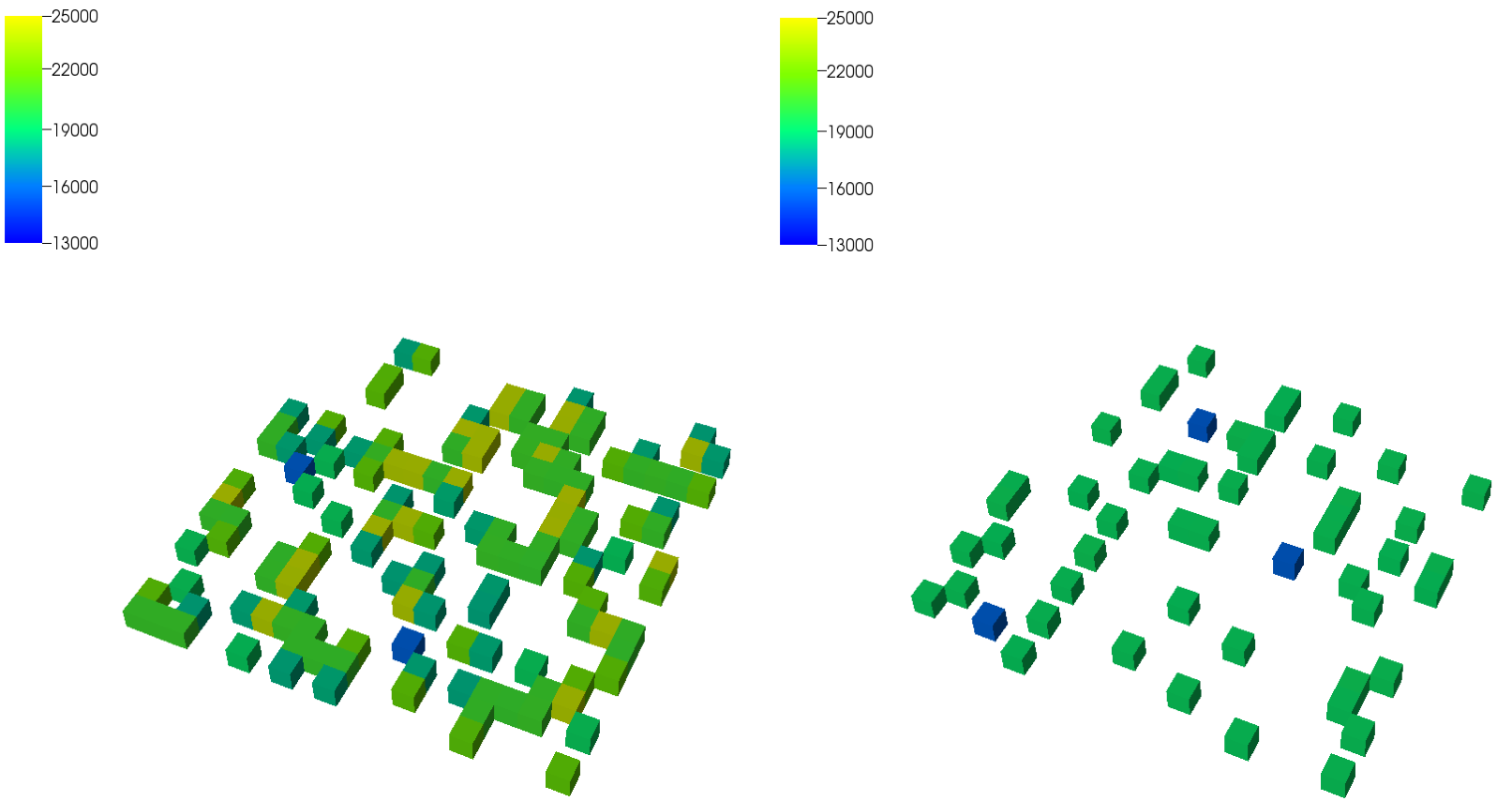}}
\caption{{Effect of water cluster on surface coverages. On the left, surface coverages (H$_2$O~nm$^{-2}$) as a function of time (s) at $n_{\rm H_2O}\,=\,6\,\times\,10^{10}$~cm$^{-3}$ and $T\,=\,700$~K, with cluster effect (pink curve) and without (light blue curve). On the right, "snapshots" of the grid resembling the [100] forsterite surface showing the adsorption energy of the occupied sites without (middle plot) and with (right plot) cluster effect. The color scale indicates the adsorption energy in the range $13\,000$~K (blue) - $25\,000$~K (yellow).}}
\label{fig:cluster}
\end{figure*}

The formation of clusters is therefore a competition between the evaporation of individual water molecules and the encounter of two molecules to initiate the cluster. In our models this effect can be seen in Fig.~\ref{fig:coverage} of Sec.~3.2 and here in Fig~\ref{fig:cluster}. At equal ($T, n_{\rm H_2O}$) conditions,~$35$\% of the surface sites are covered by water molecules when they form clusters with neighboring molecules (left plot, pink curve). The binding energies of the adsorbed water increase up to~$25\,000$~K (right image) when clusters occur with four neighboring molecules. When the cluster effect is switched off, our simulations show that the surface coverage is~$40$\% reduced (left plot, blue curve) and the adsorbed water molecules possess on average~$20\,000$~K binding energy with the surface (middle plot).
\end{appendix}
\end{document}